\documentclass[twocolumn,aps,pra,groupedaddress]{revtex4-1}
\pdfoutput=1 
\usepackage{graphicx}
\usepackage{amssymb}
\usepackage{amsmath}
\usepackage{bm}
\usepackage{bbm} 
\usepackage{subfigure}
\usepackage{color}

\newcommand{\set}[1]{\left\{#1\right\}}

\newcommand{\ket}[1]{|{#1}\rangle}
\newcommand{\bra}[1]{\langle{#1}|}
\newcommand{\abs}[1]{\left|{#1}\right|}
\newcommand{\fun}[1]{\left(#1\right)}
\newcommand{\ival}[1]{\left[#1\right]}

\newcommand{\ee}{\mathit{e}}
\newcommand{\ii}{\mathit{i}}

\newcommand{\Tr}{\mathrm{Tr}}

\begin{document}

\title{Coherence creation in an optically thick medium by matched propagation of a chirped laser pulse pair}
\author{N. Sandor, G. Demeter, D. Dzsotjan and G. P. Djotyan }
\affiliation{Wigner Research Center for Physics, Hungarian Academy of Sciences, Konkoly-Thege Mikl\'os \'ut 29-33, H-1121 Budapest, Hungary}
\date{\today}
\begin{abstract}
We consider the simultaneous propagation of a pair of Raman-resonant, frequency-modulated (chirped) laser pulses in an optically thick medium, modeled by an ensemble of $\Lambda$-atoms. A self-organization (`matching') effect is shown for the chirped pulse pair, which leads to a quasi-lossless propagation. Furthermore, we demonstrate that a well-defined coherent superposition of the atomic ground states and, correspondingly, a coherence is robustly created in the medium that can be controlled by amplitudes of the laser pulses. The proposed scheme can be applied to substantially increase the efficiency of the optical wave mixing processes, as well as in other nonlinear processes where the initial preparation of a spatially extended medium in a coherent superposition state is required.\end{abstract}
\maketitle

\section{Introduction}

Adiabatic control (AC) of atomic quantum states (see~\cite{Bergmann1998, Vitanov2001, Kral2007} and
references therein) is a powerful technique in quantum optics that
allows various applications in nonlinear optics, such as high harmonic
generation~\cite{Watson1996}, multiphoton ionization~\cite{Jones1995},
nonlinear frequency conversion~\cite{Kis2003, Eilam2006} and several
transparency effects~\cite{Wang2000}.  The general aim of AC is to create a given population distribution
among the working levels of atomic systems in a robust way using a
limited number of laser pulses. 

One important control problem is the creation of coherence between specific states of single atoms or --- allowing more exciting applications --- ensembles of atoms. It has been shown~\cite{Scully1992191} that preparing an optically thick medium of $\Lambda$-atoms (see Fig. (\ref{fig:lambda})) in the coherent superposition of their ground states leads to numerous interesting phenomena. These include lasing without inversion~\cite{Scully1992, Mompart2000}, enhanced index of refraction of the medium~\cite{Zibrov1996, Lukin1999}, Ramsey interference~\cite{Kolesov2006} or enhancement of magneto-optical effects~\cite{Sautenkov2000, Wojciechowski2010}.

In our present paper we propose a method for achieveing this task by using a pair of frequency chirped (FC) laser pulses. We anticipate that the presented method allows a relatively precise control of the final state of the medium, and is easier to apply in certain experimental situations than other methods known in literature~\cite{Kozlov2009}. The problem of the preparation of an extended medium naturally raises the question whether a lossless propagation of a pair of FC pulses is possible in the medium, and if so, by what mechanism? In this communication, we wish to give a detailed response to these questions.

The essence of AC is adiabatically tuning one of the parameters of the atom-laser interaction in time, which drives the atomic populations along the adiabatic
states of the system~\cite{Oreg1984, Grischkowsky1976}. If the evolution takes place along the dark state (population trapping), complete population control can be achieved without excitation of the atom. This mechanism is the basis of the stimulated Raman adiabatic passage (STIRAP)~\cite{Gaubatz1990, Unanyan2001, Peters2005, Vitanov1999}. In the AC schemes based on STIRAP, two (or more) time-shifted laser pulses with constant carrier frequencies are applied to a quantum system resulting in adiabatic altering of couplings between the laser fields and the atomic transitions. 

FC laser pulses applied in the atom-laser interaction represent another possibility for performing AC. In this case, the frequency of the driving electromagnetic field(s) is the key parameter governing the rearrangement of the atomic population among its quantum states. Although perfect population trapping in ground states cannot be realized when AC is performed by FC pulses (like in the STIRAP-based schemes), numerous AC schemes were proposed with negligible atomic excitation\cite{Sola1999, Djotyan2004, Djotyan2008, Sandor2011}. 

The above mentioned works generally concentrate on AC in single atoms and neglect the back-action of the atoms on the laser fields along with other propagation effects, such as interaction of the laser pulses with each other. The latter effects, however, have to be taken into account when AC is performed in an optically thick medium~\cite{Fleischhauer1999, Demeter2007, Kozlov2009, Siddons2012}. In this case, preparation of the atoms of a medium in coherent superposition of the quantum states may significantly modify its optical properties leading to very interesting and important propagation effects.

Most works in literature deal with propagation of transform-limited laser pulse(s) (without modulation of their frequency over time). In the ---most known--- case of electromagnetically induced transparency (EIT) (see~\cite{Marangos1998, Fleischhauer2000, Wang2004} and references therein), an intense laser-pulse (of constant carrier frequency) renders the whole medium transparent for a weak probe pulse in Raman resonance with the intense one. In these schemes, no population redistribution occurs in the first order.

Nearly lossless propagation was demonstrated in the case of constant frequency pulses having identical~\cite{Harris1993} and complementary pulse envelopes~\cite{Grobe1994} in optically thick media consisting of $\Lambda$-atoms. In the above mentioned schemes, the lossless propagation of the electromagnetic field(s) was ensured by initially preparing the medium in a dark superposition of the ground states. This means that no excitation occurs in the atoms during the interaction, which significantly reduces the back-action of the atoms on the laser field. As a result, basically the same population-control mechanism was established in the atoms of the extended medium as in a single atom, even for significant propagation distances.

On the other hand, it has been shown in~\cite{Harris1995} that for a sufficiently intense laser pulse pair having constant frequencies in Raman-resonance, lossless propagation is possible in a medium of $\Lambda$-atoms even if the initial preparation of the atoms does not coincide with the dark state. The explanation is that the laser pulses become distorted by the interaction with the medium in such a way that the initial preparation of the atoms corresponds to a dark state for the pulse-pair after propagating some distance. In this sense, the interacting laser pulses become \textit{matched} to each other through the interaction with the atoms of the medium.

The propagation of FC pulses in an extended medium, however, is less studied. In our earlier works,~\cite{Djotyan2001} and~\cite{Djotyan2004}, we considered several interaction schemes including $\Lambda$-atoms and FC pulses. Comparing the schemes presented there one may conclude that the population transition process induced among the states of the $\Lambda$-atom depends on whether the coupling of the two ground states are in Raman-resonance. That is, it was shown in~\cite{Djotyan2004} that a nearly excitation-free population transfer may be established among the ground states with a single FC pulse which can couple both of the transitions, provided that there is an energy difference between them (i.e. there is a Raman-detuning between the couplings). Based on this scheme, the possibility of the quasi-lossless propagation of a single FC was shown~\cite{Demeter2007}.

The action of a pair of strong Raman-resonant FC pulses on a single atom results in adiabatic excitation of the bright component of the superposition of the ground states leaving intact the dark component of this superposition~\cite{Djotyan2001}. As a result, a coherent superposition of the ground states is robustly created along with excitation of the atom. This excitation, however, is detrimental for the created coherence: One has to transfer the population of the excited state to another ground state to preserve the created coherence from the destructive effect of the spontaneous decay.

In the present investigation, we show that the Raman-resonant FC pulse pair, which would cause large excitation in a single atom, is modified by the medium in such a way that it no longer causes any significant excitation, and induces a coherent redistribution of the populations among the ground states of the atoms of the medium. It turns out that the matched propagation of the FC pulse pair posesses an important feature which is not typical for the matched propagation of pulses with constant carrier frequencies. Namely, the FC pulse pair, while propagating in a quasi-lossless way in the medium, prepares its atoms in a well-defined superposition of the metastable ground states. The created coherent superposition can be controlled by the peak amplitudes of the interacting laser pulses. The robust creation of (maximum) coherence between the ground states of the atoms in an extended optically thick medium for applications in  nonlinear frequency conversion processes is an important motivation of this investigation.

The paper is organized as follows. In Sec.~\ref{sec:mm}, a semiclassical model is presented for describing the interaction of the classical FC pulse pair with the optically thick medium composed of $\Lambda$-atoms. We present our results in Sec.~\ref{sec:prop}. We analyze the behavior of the atomic states and the interacting laser pulse pair in a transformed (symmetric-antisymmetric) basis at the boundary and inside the medium in Subsecs~\ref{sec:bound} and~\ref{sec:inside}, respectively,  based on numerical calculations and also by using the adiabatic approximation. We discuss our results in the original basis in Subsec~\ref{sec:orig} comparing them with the case of the propagation of a constant-frequency pulse pair. Our findings are summarized in Sec.~\ref{sec:sum}.

\section{Mathematical model}
\label{sec:mm}  
We use a semiclassical approach for studying the propagation of the FC pulses in a medium of $\Lambda$-atoms: the internal electronic state of the atoms is treated in the frame of quantum mechanics, while the laser pulses are described by the classical electric field 
\begin{align}
\vec{E}\left(x,t\right)=\sum_{i=1}^2{\vec\varepsilon_i\left[\mathcal{E}_i\left(x,t\right)\cdot\ee^{-\ii \left(\omega_i t- k_i x \right)}+c.c.\right]}.\label{eq:field}
\end{align}
The propagation of this field is given by the classical Maxwell equations.  In
Eq. (\ref{eq:field}) $\mathcal{E}_i$ is the complex field amplitude of
the $\text{i}^{\text{th}}$ laser pulse, which changes slowly in time
and space compared to the frequency $\omega_i$ and wave number $k_i$
($i \in \set{1,2}$). The propagation of the pulses is considered in
one direction.  The central frequencies of the laser pulses are given by $\omega_i$. In what follows, we assume that the pulses on the input boundary of the medium have a same linear frequency modulation (chirp), the range of which is  much smaller than the transition frequencies.  This allows us to incorporate the chirp into the slowly varying complex field amplitude $\mathcal{E}_i$ as a time-dependent phase: $\mathcal{E}_i\sim\exp\fun{\ii\beta t^2}$, $\beta$ being referred to as the `speed of the chirp'. 

\begin{figure}
\includegraphics[width=6.0cm, clip=true]{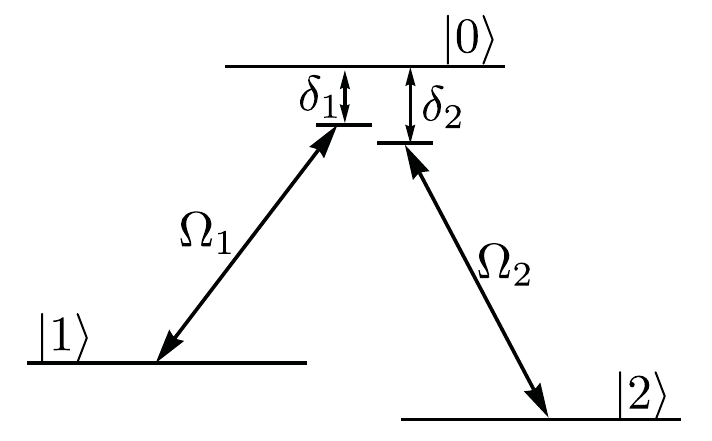}
\caption{Level scheme of the $\Lambda$-atom. The dipole-allowed
  transitions (between the excited state $\ket{0}$ and the lower
  metastable states $\ket{1}$ and $\ket{2}$, respectively) are each
  coupled by a frequency modulated (FC) laser pulse. In this communication, we consider Raman-resonant coupling, i.e. $\delta_1=\delta_2=0$.}
\label{fig:lambda}
\end{figure}

The medium in which this pair of classical FC pulses propagates  is
modeled by identical, noninteracting and motionless atoms, which are all initially prepared in one of their internal ground states (for example, in state $\ket{2}$.) We divide
the medium into small segments, inside which we can neglect the back action of the atoms on
the electric field. Since the dipole-allowed transitions between two metastable states and
a common excited state are quasi-resonantly driven, the atom can be described on the
Hilbert-space spanned by the energy-eigenstates (bare states)
$\left\{\ket{0},\ket{1}, \ket{2}\right\}$ with lambda-type linkages
(Figure \ref{fig:lambda}), where each pulse couples one transition. In our model, the difference of the transition frequencies is small, thus $\omega_1\approx\omega_2$ and $k_1=k_2=k_x$ hold for the central frequencies and wave numbers of the coupling laser pulses.
This coupling scheme may be realized by an $F=1\to F'=0$ transition in an atom interacting with two laser pulses having $\sigma^+$ and $\sigma^-$ polarizations.

The interaction of an arbitrary atom localized in the region $\left[x,x+\delta x\right]$ with the laser pulses is described in rotating wave approximation(RWA)~\cite{aleb} by the Hamiltonian
\begin{align}
\hat{H}\left(x,t\right)=-\hbar\sum_{k=1}^{2}{\left[-\delta_k\ket{k}\bra{k}+\left(\Omega_k\left(x,t\right)\ket{0}\bra{k}+\mathrm{ H.c.}\right)\right]},
\label{eqn:Ham}
\end{align}
where 
\begin{align}\Omega_k\left(x,t\right)=\frac{\mathcal{E}(x,t)d_{0k}}{\hbar}\equiv\abs{\Omega_k}\cdot\mathit{e}^{\mathit{i}\phi_k\left(x,t\right)},\enspace k\in\set{1,2}
\label{eqn:Rabi}
\end{align}
is the Rabi-frequency of the $k^\text{th}$ laser pulse, with the change of the frequency caused by the chirp included in the phase $\phi_k\left(x,t\right)$. $d_{0k}=\bra{0}\hat{d}\ket{k}$ is the matrix element $\left(0,k\right)$ of the dipole operator $\hat{d}$, whereas  $\delta_k$ is the detuning of the central frequency ($\omega_k$) of the $k^{\text{th}}$ pulse from the transition frequency between states $\ket{0}\leftrightarrow \ket{k}$. Here we consider a case where these `central detunings' are equal --- $\delta_1=\delta_2$ ---, which we now set to be zero for the sake of simplicity. Note that the frequency modulation of both pulses entering the medium follows the same time dependence at the boundary of the medium. Therefore, at the boundary of the medium, the differences between the pulses' instantenous frequencies and the corresponding atomic transition frequencies are equal in every time point, i.e. the pulses which enter the medium are in `Raman-resonance' with the atoms.

For later convenience it is worth introducing $\xi=x/\xi_0$ and $\tau=\left(t-x/c\right)/\tau_{\sigma}$ dimensionless, space and retarded time coordinates. The time is measured in the unit of $\tau_\sigma$, which characterizes the duration of the pulses. For the normalization of the space coordinates we introduce the absorption length of a laser pulse of constant frequency $\omega_L$ in a medium consisting of  resonant two-level atoms with a density of $\mathcal{N}$, which is given by~\cite{aleb} 
\begin{align}
\xi_{0}=\frac{\varepsilon_0\hbar c}{\mathcal{N}\omega_L\abs{d_A}^2T},
\end{align}  
where $T$ is the natural lifetime of the excited state and $d_A$ is the dipole momentum of the coupled atomic transition. Although there are two atomic transitions in the
present case, it is consistent with our previous approximations to regard a common absorption length for both coupling pulses by setting $\abs{d_{01}}=\abs{d_{02}}=d_A$, $\omega_L=\left(\omega_{1}+\omega_2\right)/2$ and $T=2/\Gamma_1=2/\Gamma_2=2/\Gamma$, where $\Gamma_i$ is the longitudinal relaxation rate from the excited state $\ket{0}$ to the metastable state $\ket{i}$, $i\in\set{1,2}$.

We describe the response of the atoms for the ingoing laser radiation inside a certain space-interval by the master equation
\begin{align}
&\nonumber\partial_\tau\hat{\rho}\left(\xi,\tau\right)=\frac{1}{\ii\hbar} \left[\hat{H}\left(\xi,\tau\right),\hat{\rho}\left(\xi,\tau\right)\right]- 2\Gamma\ket{0}\bra{0}\rho_{00}\left(\xi,\tau\right)\\\nonumber&\quad+ \sum_{k=1}^2{\left[\Gamma\rho_{00}\left(\xi,\tau\right)\ket{k}\bra{k} - \left(\Gamma\rho_{0k}\left(\xi,\tau\right)\ket{0}\bra{k}+\mathrm{ H.c.}\right)\right]}\\
&\qquad \left(\rho_{kl}=\bra{k}\hat{\rho}\ket{l} \right),
\label{eq:master}
\end{align}
where the density matrix as a function of the space coordinate $\xi$ is defined by the average 
\begin{align}
\hat{\rho}\left(\xi,\tau\right)=\frac{1}{N} \sum_{i=1}^N {\hat\rho^{(i)}\left(\tau\right)}.
\end{align} 
Here $N$ is the number of the atoms inside the space interval $\left[\xi,\xi+\delta\xi\right]$ and $\rho^{\left(i\right)}$ denotes the density matrix of the $i^{\text{th}}$ atom. Since the change in the electromagnetic field is neglected inside a small segment, the Hamiltonian $\hat{H}\left(\xi,\tau\right)$ which drives the evolution of the average density operator is formally the same as in Eq. (\ref{eqn:Ham}).

The atoms of the medium may affect the propagating laser pulses by means of basically two mechanisms: by spontaneous emission from the excited state and by dipole-radiation~\cite{Harris1995}. Here we put emphasis on the latter by regarding a weakly decaying limit, where the lifetime of the excited state is assumed to be about an order of magnitude longer than the interaction time. The macroscopic polarization induced in the medium by the laser fields may be written as:
\begin{align}
\vec{P}\left(\xi,\tau\right)=N\Tr{\hat{\rho}\left(\xi,\tau\right)\hat{d}}= N\sum_{k=1}^2{\left[\rho_{0k}\left(\xi,\tau\right)d_{0k}+\mathrm{ H.c.}\right]}.\label{eqn:macpol}
\end{align}

This quantity serves as a source in the Maxwell equation which is used to describe the dynamics of the electric field. In consistence with the RWA, one can apply the slowly varying envelope approximation~\cite{Duarte1997} in order to get a first order differential equation for the propagation from the second- order wave-equation. Using these approximations, one gets the following differential equations for the Rabi-frequencies of the laser pulses(see Eq. (\ref{eqn:Rabi})):
\begin{align}
\frac{\partial}{\partial\xi}\Omega_{k}\left(\xi,\tau\right)=-\mathit{i}\alpha\rho_{k0}\left(\xi,\tau\right),\enspace k\in\set{1,2}
\label{eqn:Max_Rabi}
\end{align}
where $\alpha=\tau_{\sigma}/\fun{2 T}$ describes the strength of the coupling between the medium and the lasers. 

Eqs. (\ref{eq:master}) and (\ref{eqn:Max_Rabi}) together form a system of partial differential equations, which describes the coupled laser-atom system in the extended medium. 
In order to study the behavior of the propagating FC pulses and the atomic transitions induced by the laser field, we solve this system numerically. We use the following boundary conditions of two Gaussian, linearly chirped pulses entering the medium at $\xi=0$ (see Fig~\ref{fig:Ica}):
\begin{subequations}
\begin{align}
&\begin{bmatrix}\Omega_{1}\fun{\xi=0,\tau}\\\Omega_{2}\fun{\xi=0,\tau}\end{bmatrix}=\begin{bmatrix}\vartheta_1\\\vartheta_2\end{bmatrix}e^{-\tau^2\fun{\frac{1}{2}+\ii\beta}},\\\nonumber&\vartheta_k\in\mathbb{R}\enspace\forall k\in\set{1,2}\\
\nonumber\\
&\hat{\rho}\left(\xi,\tau\to-\infty\right)=\ket{2}\bra{2},
\end{align}
\label{eqn:bc}
\end{subequations}
where $\vartheta_1$ and $\vartheta_2$ are the peak amplitudes of the laser pulses at the entrance of the medium, and $\beta$ is referred to as the `speed of chirp'. The parameters $\set{\vartheta_1,\vartheta_2,\beta,\tau_{\sigma}}$ in what follows are chosen in such a way that the conditions of adiabaticity~\cite{Bergmann1998} are fulfilled.

\subsection{Symmetric-antisymmetric basis\label{subsec:basis}}
For further investigation of the population dynamics in arbitrary segments of the medium, let us introduce a basis transformation on the atomic states and the electric field modes adapted to the boundary conditions given in Eq.~\eqref{eqn:bc}. This transformation leads us to the following symmetric-antisymmetric basis and effective Rabi frequencies:
\begin{subequations}
\begin{align}
&\set{\ket{0},\ket{s},\ket{a}}=\set{\ket{0},\frac{\vartheta_1\ket{1}+\vartheta_2\ket{2}}{\vartheta}, \frac{\vartheta_2\ket{1}-\vartheta_1\ket{2}}{\vartheta}},\\
&\begin{bmatrix}\Omega_s\\\Omega_a\end{bmatrix}=\begin{bmatrix}\left(\vartheta_1\Omega_1+\vartheta_2\Omega_2\right)/\vartheta\\ \left(\vartheta_2\Omega_1-\vartheta_1\Omega_2\right)/\vartheta\end{bmatrix},\label{eqn:transRabis}
\end{align}
\end{subequations}
where $\vartheta=\sqrt{\vartheta_1^2+\vartheta_2^2}$, and $\Omega_s$ and $\Omega_a$ give the couplings between the excited state $\ket{0}$ and the symmetric and antisymmetric superpositional states $\ket{s}$ and $\ket{a}$, respectively.
The Hamiltonian in this new basis becomes
\begin{align}
\hat{H}_{sa}\fun{\xi,\tau}&=-\hbar\sum_{j\in{s,a}}{\ival{\Omega_j\fun{\xi,\tau}\ket{0}\bra{j}+\mathrm{H.c.}}}.\label{eqn:Hamsa}
\end{align}
The dynamics of the effective Rabi frequencies formally obeys the same differential equation as the original ones:
\begin{align}
\partial_\xi\Omega_j&=-\ii\alpha\rho_{j0},\enspace j\in\set{s,a}, \rho_{j0}=\left\langle j\mid\hat{\rho}\mid 0\right \rangle
\label{eq:max_sa}
\end{align}
where $\vartheta_k\in\mathbb{R}\enspace\forall k\in\set{1,2}$ was utilized.
\begin{figure}[!hbt]
\subfigure[\label{fig:Ica}]{\resizebox{4.2 cm}{!}{\includegraphics*{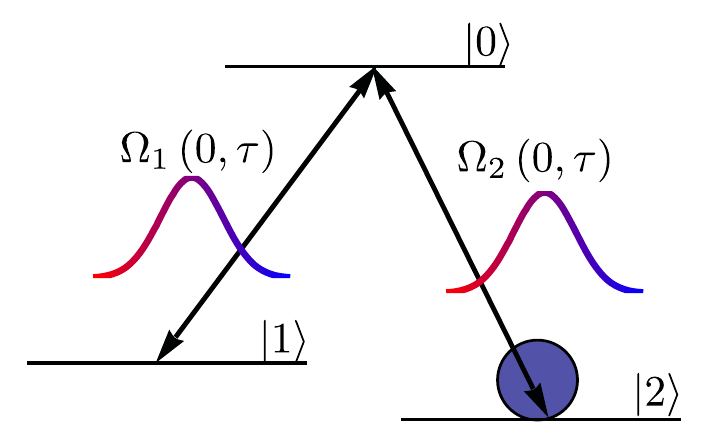}}}
\subfigure[\label{fig:Icb}]{\resizebox{4.2 cm}{!}{\includegraphics*{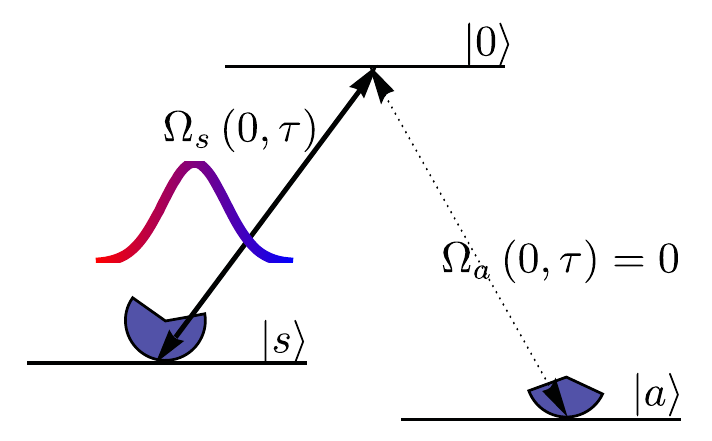}}}
\caption{(Color online) \textbf{Boundary conditions given in a.) the original atomic basis b.) the symmetric-antisymmetric basis.} All the atoms in the medium are initially prepared in the metastable state $\ket{2}$, which corresponds to a coherent superposition in the symmetric-antisymmetric basis given by the peak amplitudes $\vartheta_1$ and $\vartheta_2$ of the ingoing laser pulses.}
\label{fig:Ic}
\end{figure}
Transforming the boundary conditions according to Eq.~\eqref{eqn:transRabis} yields a set of boundary conditions which is more convenient for our purposes, as it only contains one ingoing laser mode (see Fig~\ref{fig:Icb}):

\begin{subequations}
\begin{align}
&\begin{bmatrix}\Omega_{s}\fun{\xi=0,\tau}\\ \Omega_{a}\fun{\xi=0,\tau}\end{bmatrix} =\begin{bmatrix}\vartheta\\0\end{bmatrix}e^{-\tau^2\fun{\frac{1}{2}+\ii\beta}},\\
&\hat{\rho}\left(\xi,\tau\to-\infty\right)=\frac{1}{\vartheta^2}\fun{\vartheta_2\ket{s}-\vartheta_1\ket{a}}\fun{\vartheta_2\bra{s}-\vartheta_1\bra{a}}.\label{eqn:bcsa_2}
\end{align}
\label{eqn:bcsa}
\end{subequations}

\subsection{Rotating basis for adiabatic approximation}
In case of adiabatic evolution, the dynamics of the system can be well described by the analysis of the eigenstates of the interaction Hamiltonian. This is based on the fact that the evolution is slow (compared to the characteristic frequency defined by the inverse of the difference of the eigenenergies), thus the system follows the eigenstate in which it was initially prepared in~\cite{Oreg1984, Grischkowsky1976}. We can only apply this approximation if the interaction Hamiltonian contains matrix elements that vary slowly in time (compared to the timescale of the interaction). For applying this approximation, it is important to use a basis in which the Hamiltonian of the system contains slowly varying matrix elements (in the scale of the interaction time). 

We use an interaction picture for describing the atomic dynamics in each segment, applying the following rotating basis vectors:
\begin{align}
\set{\ket{0}, \ket{\tilde{s}}, \ket{\tilde{a}}}\equiv\set{\ket{0}, \ket{s}\ee^{-\ii\phi_s\fun{\xi,\tau}}, \ket{a}\ee^{-\ii\phi_a\fun{\xi,\tau}}}.\label{eqn:rotbasis}
\end{align}
As a result, the time evolution is described in each segment by the following --- slowly varying --- Hamiltonian:
\begin{align}
\nonumber\hat{\mathcal H}_{sa}\fun{\xi,\tau}=-\hbar\sum_{j\in\set{s,a}}&\left[{\partial_{\tau}\phi_j\fun{\xi,\tau}\ket{\tilde j}\bra{\tilde j}+}\right.
\\&\left.+\fun{\abs{\Omega_j\fun{\xi,\tau}}\ket{0}\bra{\tilde{j}}+\mathrm{H.c.}}\right],
\label{eqn:Hamrotsa}
\end{align}
$\phi_j\fun{\xi,\tau},\enspace j\in\set{s,a}$ being the phase of the Rabi-frequency $\Omega_j\fun{\xi,\tau},\enspace j\in\set{s,a}$ (c.f. Eq.\eqref{eqn:Rabi}). Note that this transformation does not change the absolute values of the coefficients in the system's state vector.

\section{Propagation of the FC pulse pair in the optically thick medium}
\label{sec:prop}
In order to describe the collective behavior of the atom-laser system, we  solve the system of differential equations \eqref{eq:master} and \eqref{eqn:Max_Rabi} numerically. As it was already mentioned, the propagation equations for the Rabi frequencies are the same in both, the original atomic and the transformed symmetric-anti-symmetric basis except for the boundary conditions. Thus, it is easy to analyze the system in both coordinate systems. This is beneficial because the processes can be understood better in the transformed system, whereas it is important to have the results in the atomic basis where a possible measurement could be conducted.

The ingoing coupling fields that we consider are relatively strong, having a few $10\pi$ for pulse area of the Rabi-frequencies $\Omega_1$ and $\Omega_2$. The frequency modulation is chosen so that the pulses drive an adiabatic population transfer at the boundary of the medium, with a speed of chirp  of $\beta=7 \ival{1/\tau_{\sigma}^2}$. We would like to analyze a coherent population-transfer process in the medium, so the ingoing fields are considered to be short compared to the lifetime of the excited state $T_{\ket{0}}=50 \tau_{\sigma}$. For example, for a $\mathrm{Rb}^{87}$ atom having a lifetime of $\tau=26.235 \mathrm{ns}$, the pulse length should be approximately $2\mathrm{ns}$, which means that a linear chirp speed of $28\mathrm{GHz}/\mathrm{ns}$ and maximum Rabi frequencies in the order of $30\mathrm{GHz}$ would be needed.

\subsection{Population dynamics induced by the FC pulse pair at the boundary of the medium ($\xi=0$) \label{sec:bound}}
\begin{figure}[!hbt]
\includegraphics[width=6.8cm, clip=true]{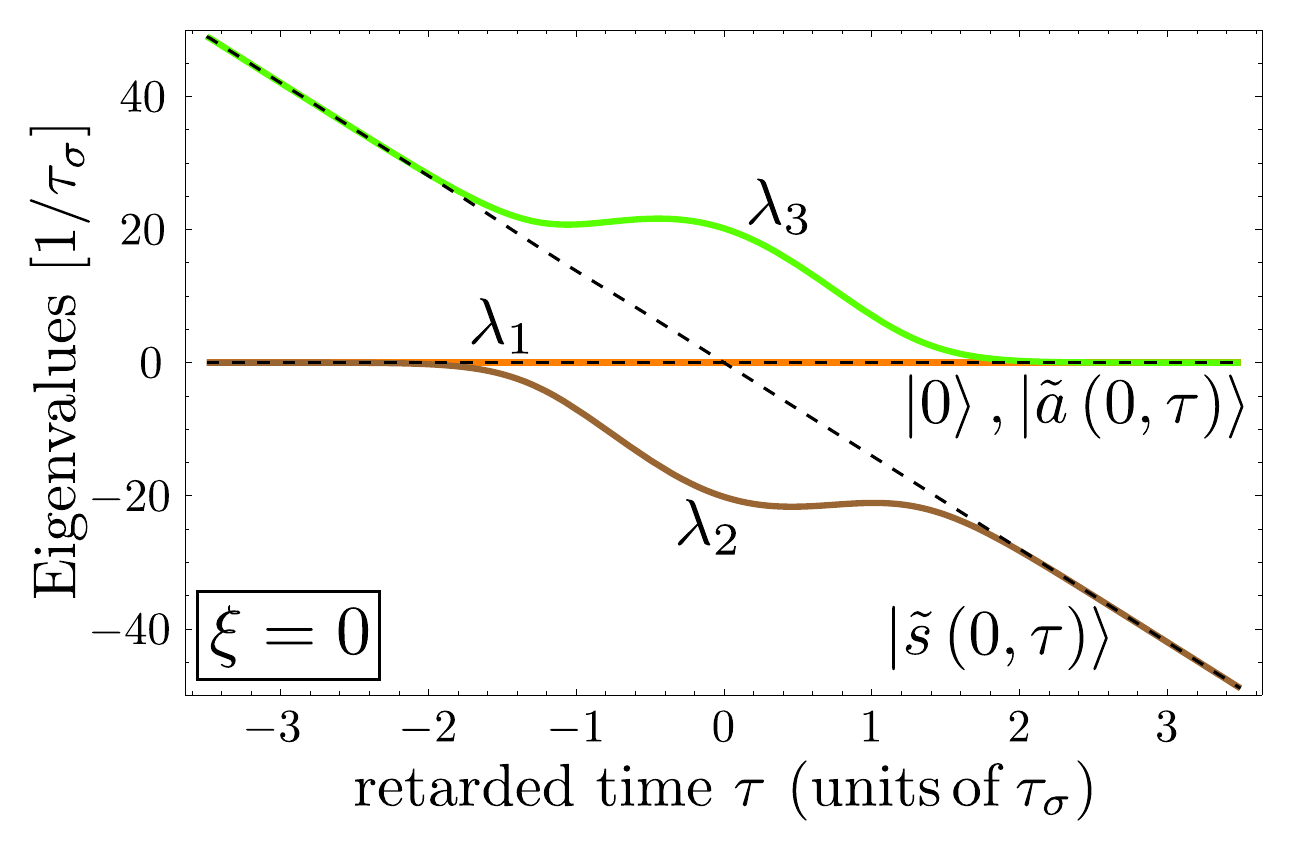}
\caption{(Color online) \textbf{Eigenvalues of the Hamiltonian $\hat{\mathcal H}_{sa}$ (Eq. \eqref{eqn:Hamrotsa}) which describes the atom-laser system at the boundary of the medium ($\mathbf{\xi=0}$).} The eigenvalues belonging to the diabatic states are plotted with dashed lines which correspond to states $\ket{0}$, $\ket{\tilde s}$ and $\ket{\tilde a}$, respectively. The eigenvalues of the adiabatic states are plotted with solid lines. The antisymmetric state $\ket{\tilde a}$ is an eigenstate of the Hamiltonian $\hat{\mathcal H}_{sa}$ with an eigenvalue of $\lambda_1=0$. The eigenstate belonging to $\lambda_2$ evolves from the excited state $\ket{0}$ to the symmetric state $\ket{\tilde s}$, while the other eigenstate belonging to $\lambda_3$ follows the inverse path ($\ket{\tilde s}\to\ket{0}$).}
\label{fig:eiv0}
\end{figure}

We first describe the interaction of the atoms with the Raman-resonant FC pulse pair at the boundary of the medium in the symmetric-antisymmetric basis. (The results are also valid approximately for an optically dilute medium). It is easy to see from Eqs.~\eqref{eqn:Hamrotsa} and \eqref{eqn:bcsa} that the antisymmetric state $\ket{\tilde a}$ is an eigenstate of the Hamiltonian $\hat{\mathcal H}_{sa}\fun{\xi=0,\tau}$ since the coupling $\Omega_a$ is 0 for $\xi=0$. The other metastable state $\ket{\tilde s}$ and the excited state $\ket{0}$ form a two-state atom coupled by a FC pulse which drives a rapid adiabatic passage from state $\ket{\tilde s}$ to the excited state $\ket{0}$,~\cite{Vitanov2001} (see Fig.~\ref{fig:eiv0}). Since the antisymmetric state is decoupled from the excited state, it represents a dark state for the Raman-resonant FC pulses. 

\begin{figure}[!hbt]
\includegraphics[width=6.8cm, clip=true]{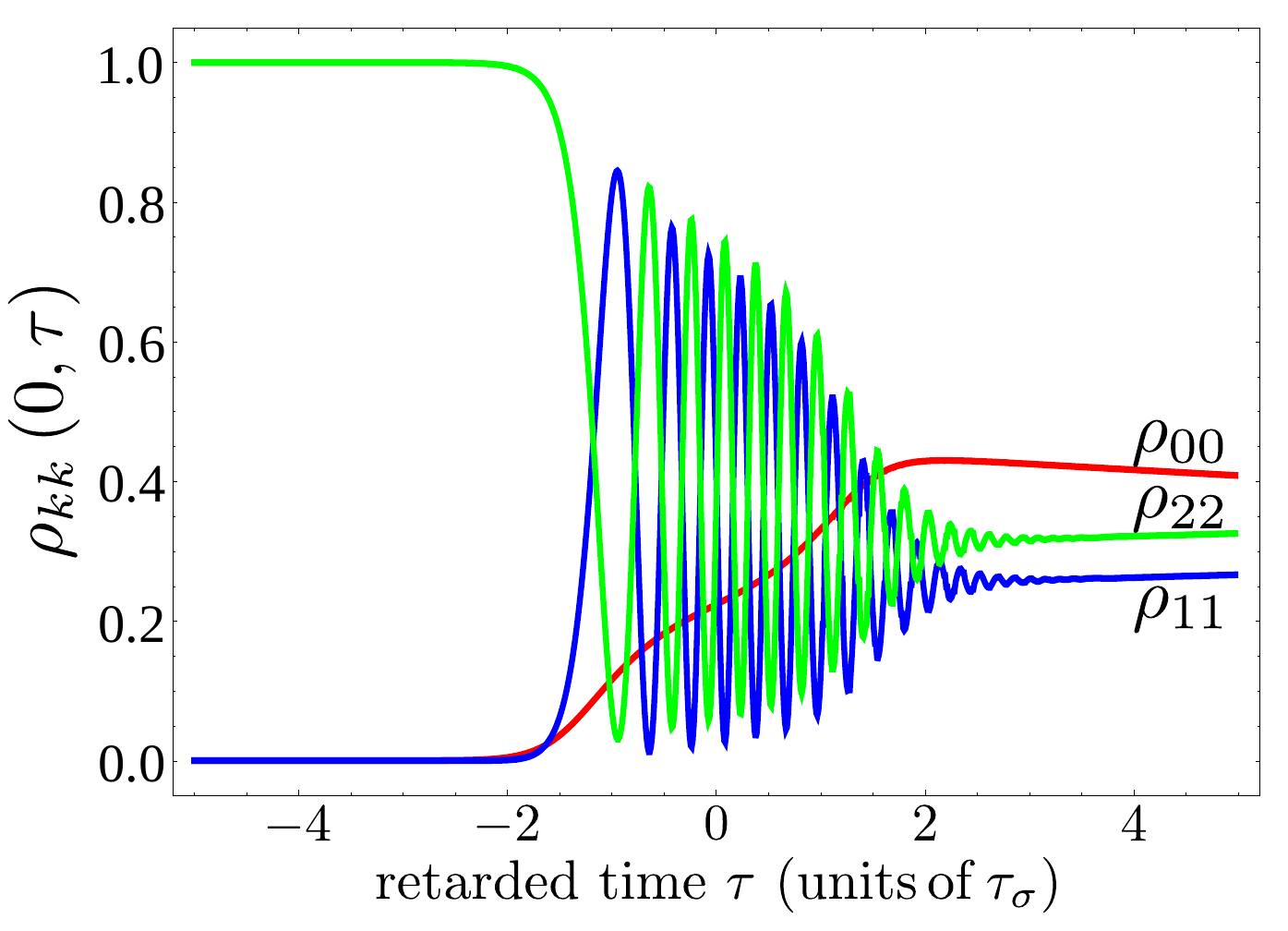}
\caption{(Color online) \textbf{Dynamics of the atomic populations in the symmetric-antisymmetric basis at the boundary of the medium} The parameters used for the calculation are $\vartheta_1=15 \ival{1/{\tau_{\sigma}}}$, $\vartheta_2=13.5\ival{1/{\tau_{\sigma}}}$, $\beta=7\ival{1/{\tau_{\sigma}^2}}$.}
\label{fig:states0}
\end{figure}

Similarly to the original case of constant frequency matched pulses propagating in the medium of $\Lambda$-atoms~\cite{Harris1994, Harris1995}, the initial preparation of the atoms given in Eq.~\eqref{eqn:bcsa} does not coincide with the "dark" state. Based on the above considerations in the frame of the adiabatic approximation, the atoms at the boundary are expected to be transferred from a superposition of the metastable states $\ket{\tilde{s}}$ and $\ket{\tilde{a}}$ to a superposition of states $\ket{\tilde{s}}$ and the excited state $\ket{0}$. This result is in perfect agreement with the numerical solution of the master Eq.~\eqref{eq:master} at $\xi=0$, which is depicted in  Fig.~\ref{fig:states0}. 

\begin{figure}[!hbt]
\subfigure[\label{fig:cohs0a}]{\resizebox{6.8cm}{!}{\includegraphics*{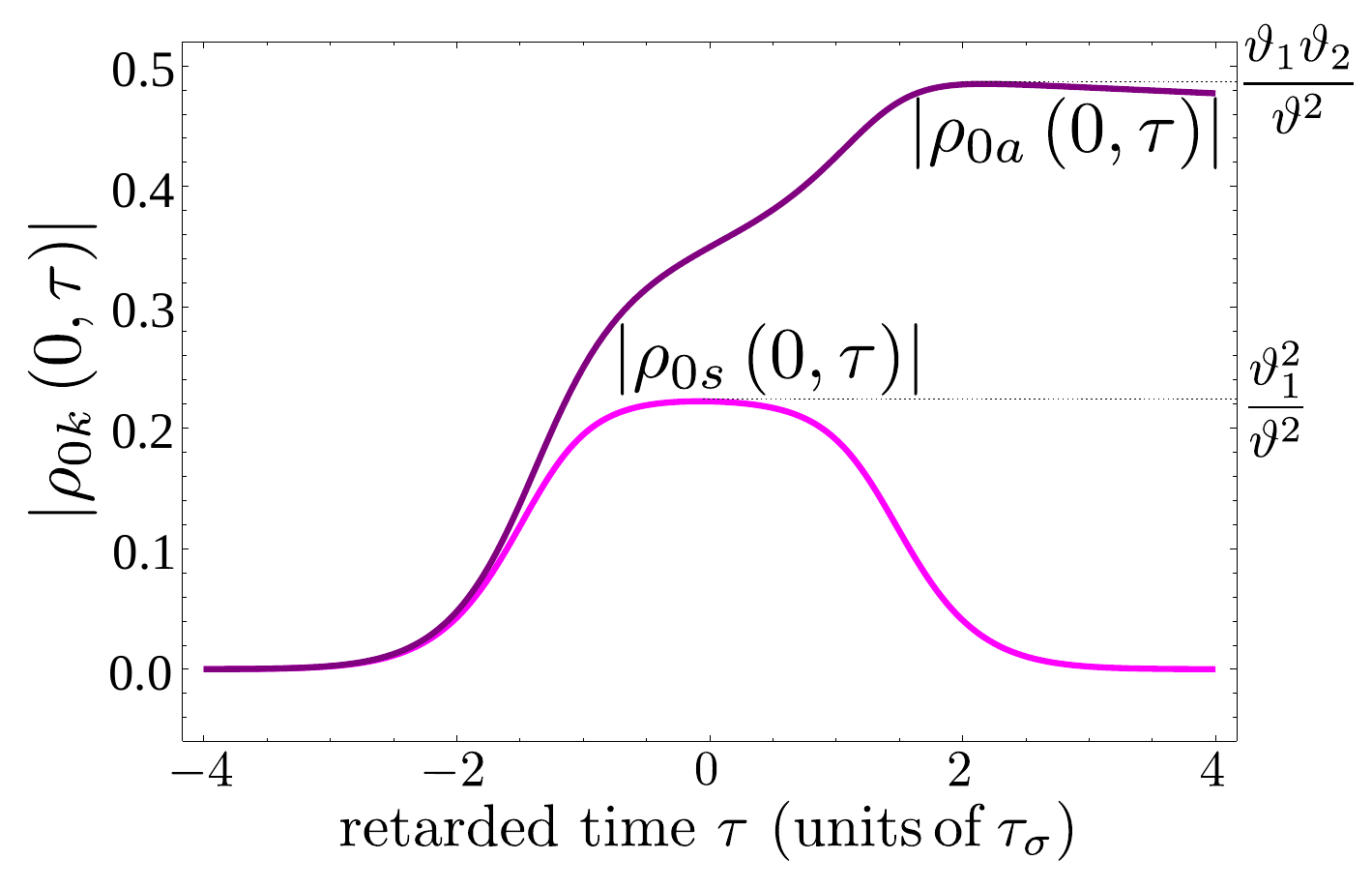}}}
\subfigure[\label{fig:cohs0b}]{\resizebox{6.8cm}{!}{\includegraphics*{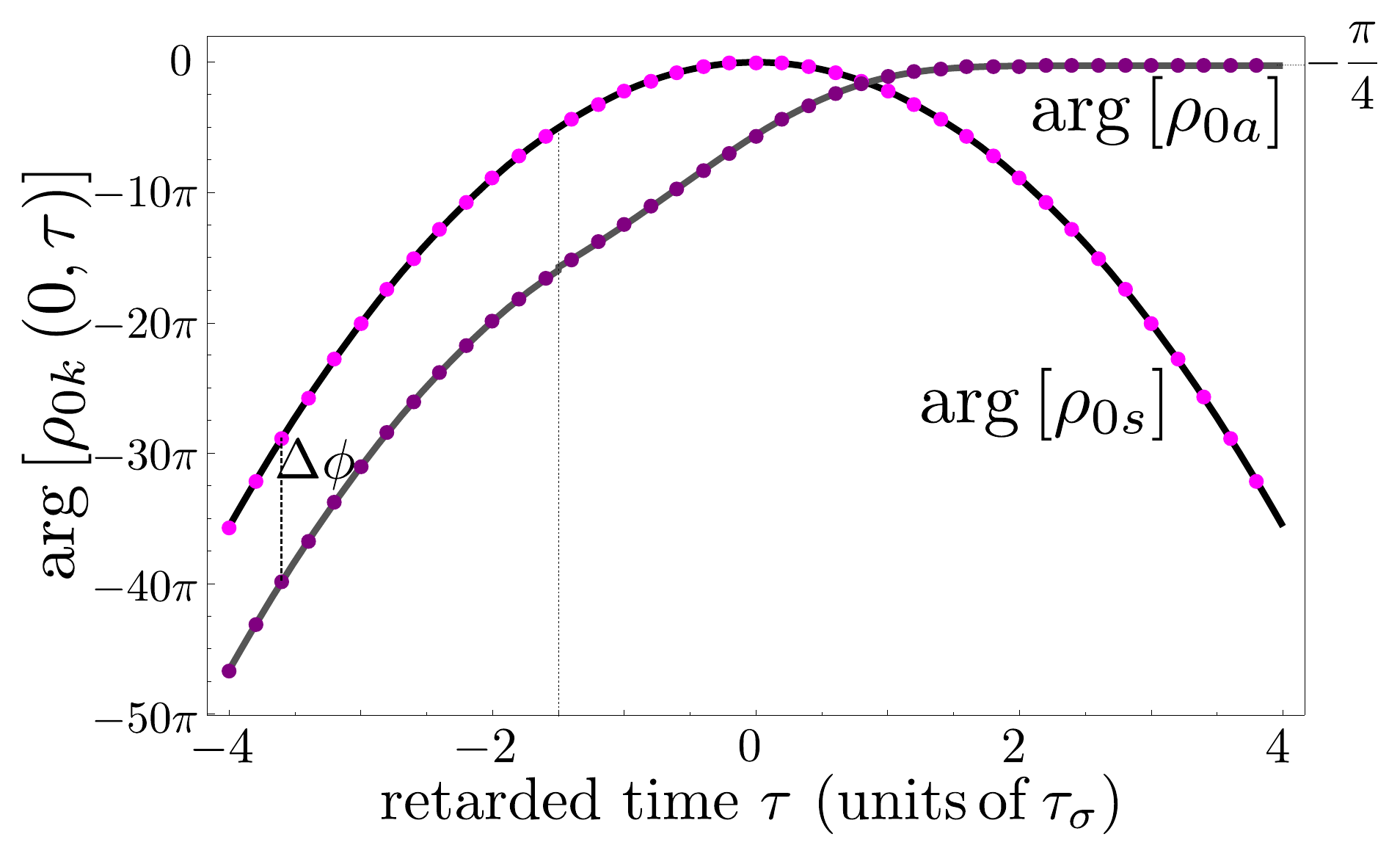}}}
\caption{(Color online) \textbf{Time evolution of a.) the absolute value and b.) the phase of the atomic coherences at the boundary of the medium.} For the absolute values of the coherences, the solid lines show the result of the numerical solution of the master equation for $\xi=0$ using the same parameters that are given in Fig.~\ref{fig:states0}. In the case of the phase of the coherences, the dots represent the result of the numerical solution, and the solid lines are the fitted functions: The function $\beta_0+\beta_1\tau^2$ is fitted on the values of $\rho_{0s}\fun{\xi=0,\tau}$ and on $\rho_{0a}\fun{\xi=0,\tau<-1.5\tau_{\sigma}}$ with the fitting parameters of $\set{\beta_1^{s}=-6.99,\beta_0^{s}=0.04}$ and $\set{\beta_1^{a}=-7.01,\beta_0^{a}=-34.35}$, with $\beta_0^{s}-\beta_0^{a}=\Delta\phi=10\pi+0.95\pi$. A different function determines the time evolution of $\rho_{0a}\fun{\xi=0,\tau>-1.5\tau_{\sigma}}$: the function $\alpha_0+\alpha_1\exp\fun{-\pi\exp\ival{\alpha_2 \tau}/2}$ is fitted on the values, with the parameters $\set{\alpha_0^{a0}=-0.78, \alpha_1^{a0}=-80.12, \alpha_2^{a0}=0.76}$}
\label{fig:cohs0}
\end{figure} 

In order to understand the laser propagation, it is important to analyze the time evolution of the coherences $\rho_{0a}\fun{\xi=0,\tau}$ and $\rho_{0s}\fun{\xi=0,\tau}$. We can easily understand the behavior of the absolute values of the coherences (see Fig.~\ref{fig:cohs0a})  from the predictions of the adiabatic approximation. Since there is a complete population transfer between states $\ket{s}$ and $\ket{0}$, the absolute value of the coherence between them is 0 in the beginning and  at the end of the interaction, and only differs from zero during the population transfer with a maximum value of  $\vartheta_2^2/\fun{2\vartheta^2}$. The time evolution of $\left|\rho_{0a}\right|$ is determined by the change of the population of the excited state $\ket{0}$ in time, since the population of state $\ket{a}$ does not change, as it is a dark eigenstate of the dressed atoms at the boundary of the medium. In parallel with the excitation of the atom, a coherence $\rho_{0a}=\vartheta_1\vartheta_2/\vartheta^2$ is established as a result of the interaction.

The time evolution of the phase of $\rho_{0s}$ is determined by the phase of the coupling pulse $\Omega_s$, which is set to be linearly chirped. It is clearly seen from Fig~\ref{fig:cohs0b} that the quadratic function $\beta_0+\beta_1\tau^2$ accurately fits the numerically calculated results of $\arg\fun{\rho_{0s}}$. The behavior of the phase of the coherence between the excited and the antisymmetric ground state $\rho_{0a}$ is more complicated. Its time function starts as the same quadratic one as for $\rho_{0s}$, but the evolution changes approximately at $\tau=-1.5\tau_{\sigma}$, and it tends to a constant value as $\alpha_0+\alpha_1\exp\fun{-\pi\exp\ival{\alpha_2 \tau}/2}$.

\subsection{Interaction of the propagating pulse pair and the atom inside the medium ($\xi>\xi_0$)}
\label{sec:inside}

\subsubsection{Laser field inside the medium}
Let us now consider the time and space evolution of the effective Rabi-frequencies $\Omega_s$ and $\Omega_a$. In the symmetric-antisymmetric basis, only one, strong coupling field ($\Omega_s$) enters the medium of $\Lambda$-atoms, which are prepared in a superposition of states $\ket{a}$ and $\ket{s}$ (c.f. Eq. \eqref{eqn:bcsa}). In the course of the coherent transition process between the atomic states induced by $\Omega_s$, a coherence is established between the antisymmetric and the excited state of the atoms close to the boundary, with a time function of its phase described in the previous subsection (c.f. Fig~\ref{fig:cohs0}). This coherence generates the laser field $\Omega_a$.

\begin{figure}[!hbt]
\subfigure[\label{fig:Rabis}]{\resizebox{6.8 cm}{!}{\includegraphics*{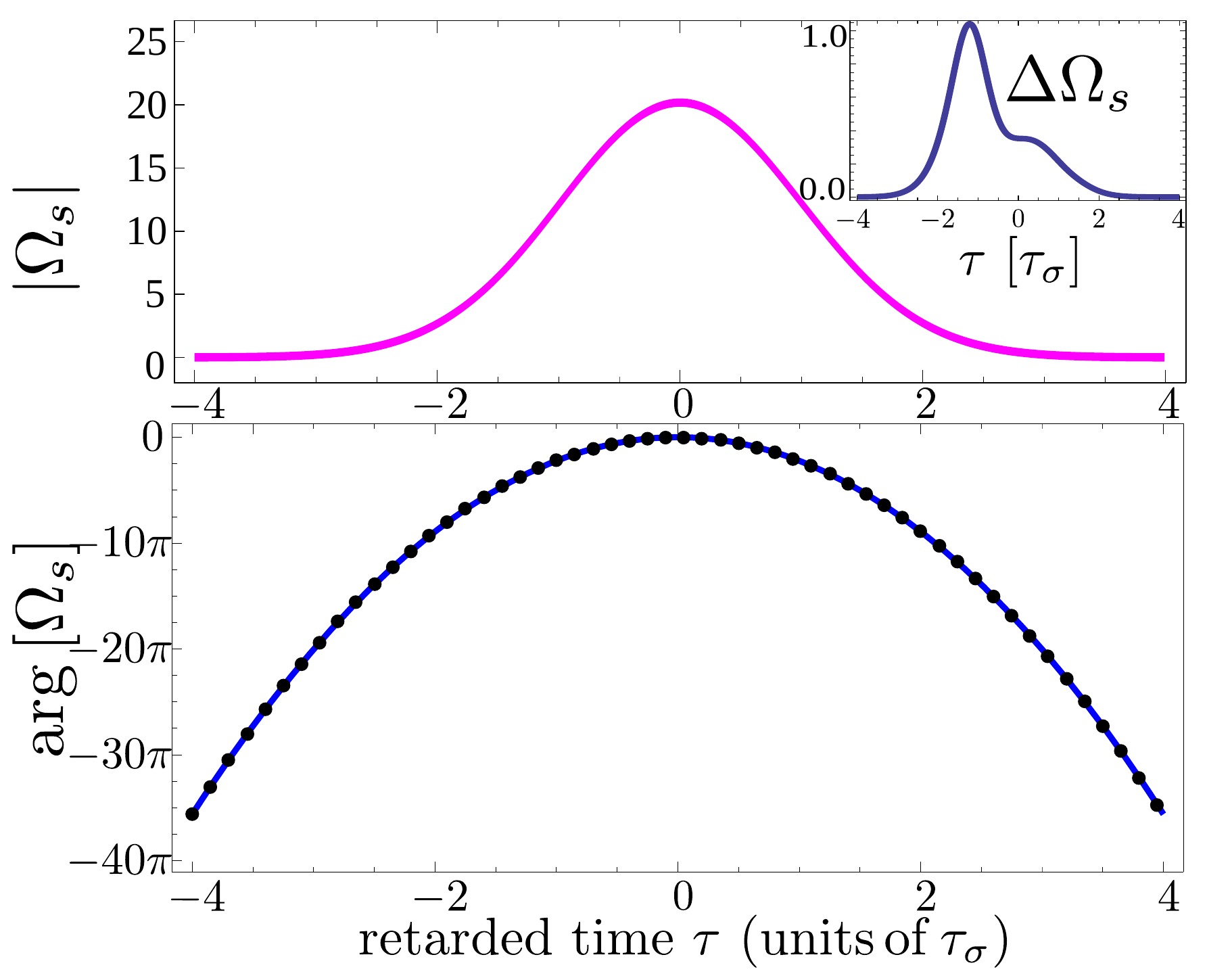}}}
\subfigure[\label{fig:Rabia}]{\resizebox{6.8 cm}{!}{\includegraphics*{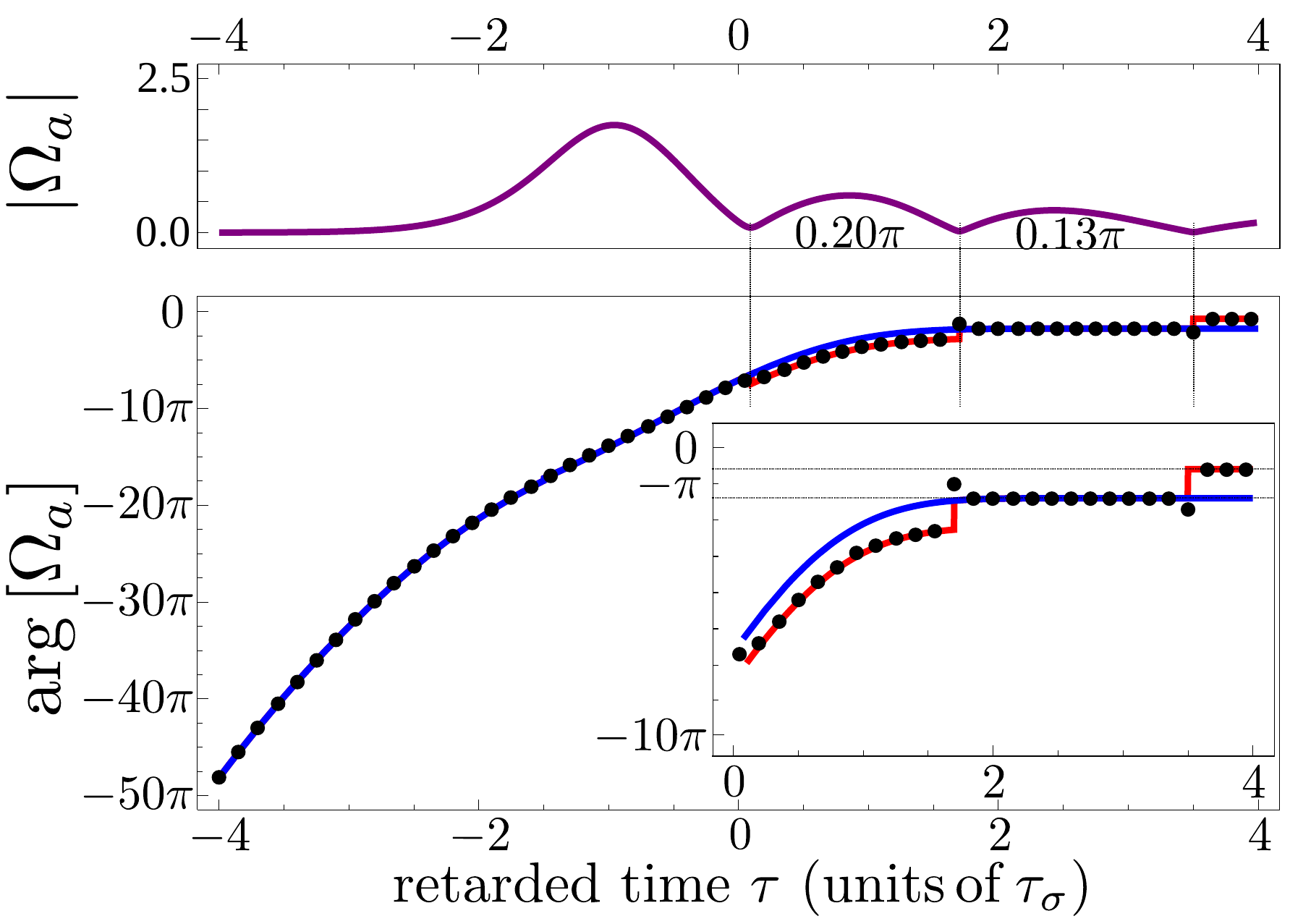}}}
\caption{(Color online) \textbf{Absolute values and phases of the effective Rabi-frequencies $\Omega_{s}$ (a.) and $\Omega_{a}$ (b.) as a function of the retarded time at a fixed space point inside the medium ($\xi=40\xi_{0}$). (Inset: Magnification of the plot showing the time evolution of the phases between $\tau=-1$ and $\tau=4$.)} The parameters used for the calculation are the same as given in Fig.~\ref{fig:states0}. The solid lines in case of the absolute values and the dots in case of the phases are results of the numerical calculation. The same function --- solid lines connecting the dots --- fits on the calculations as the one that describes the phase of $\rho_{0a}\fun{\xi=0,\tau}$ (see Fig.~\ref{fig:cohs0}) apart from 'jumps` of $\pi/2$ which correspond to the Rabi frequencies change their sign value from positive to negative, or the other way around.}
\label{fig:Rabi}
\end{figure}

The time evolution of the absolute values and phases of the effective Rabi-frequencies $\Omega_a$ and $\Omega_s$ at a given location inside the medium ($\xi=40\xi_0$) is presented in Fig.~\ref{fig:Rabi}. The coupling field $\Omega_s$ is only slightly modified during the propagation. The pulse envelope remains Gaussian  with a good approximation after a propagation of multiple times of the absorption length $\xi_0$. The change in the envelope of the effective Rabi-frequency $\Omega_s$ is presented in the inset of Fig.~\ref{fig:Rabis}. It can be observed that after $40\xi_0$ of propagation, the distortion from the boundary condition is less than $4\%$ of the pulse area. The phase function with respect to time inside the medium also has the same character as at the boundary: the same quadratic function $\beta_0+\beta_1\tau^2$ fits the data in Fig.~\ref{fig:Rabis}, so the frequency of this effective field changes linearly in time. Thus, the ingoing laser mode $\Omega_s$ preserves its initial properties during the propagation in the medium, with only a small loss in the pulse envelope.

However, the loss in the ingoing laser mode $\Omega_s$ allows for generating the another mode $\Omega_a$. The numerical calculations (see Fig.~\ref{fig:Rabia}) show that the character of the phase function of $\Omega_a\fun{\xi,\tau}$ over time is at every location very similar to $\arg(\rho_{0a}\fun{\xi=0, \tau})$, that is the  phase of the coherence between the symmetric and excited state over time.  To be more specific, the same curve can be fit onto the numerical values of $\Omega_a\fun{\xi,\tau}$ $\forall \xi$ as in case of the coherence $\rho_{0a}$ at the boundary, with almost the same fitting parameters. The only differences are a shift of $-3\pi/2$ in the constant parameters $\alpha_0$ and $\beta_0$ and a phase-jump of $\pi$ at every retarded time point where $\Omega_a=0$. 

\begin{figure}[!hbt]
\subfigure[\label{fig:cohs}]{\resizebox{6.8 cm}{!}{\includegraphics*{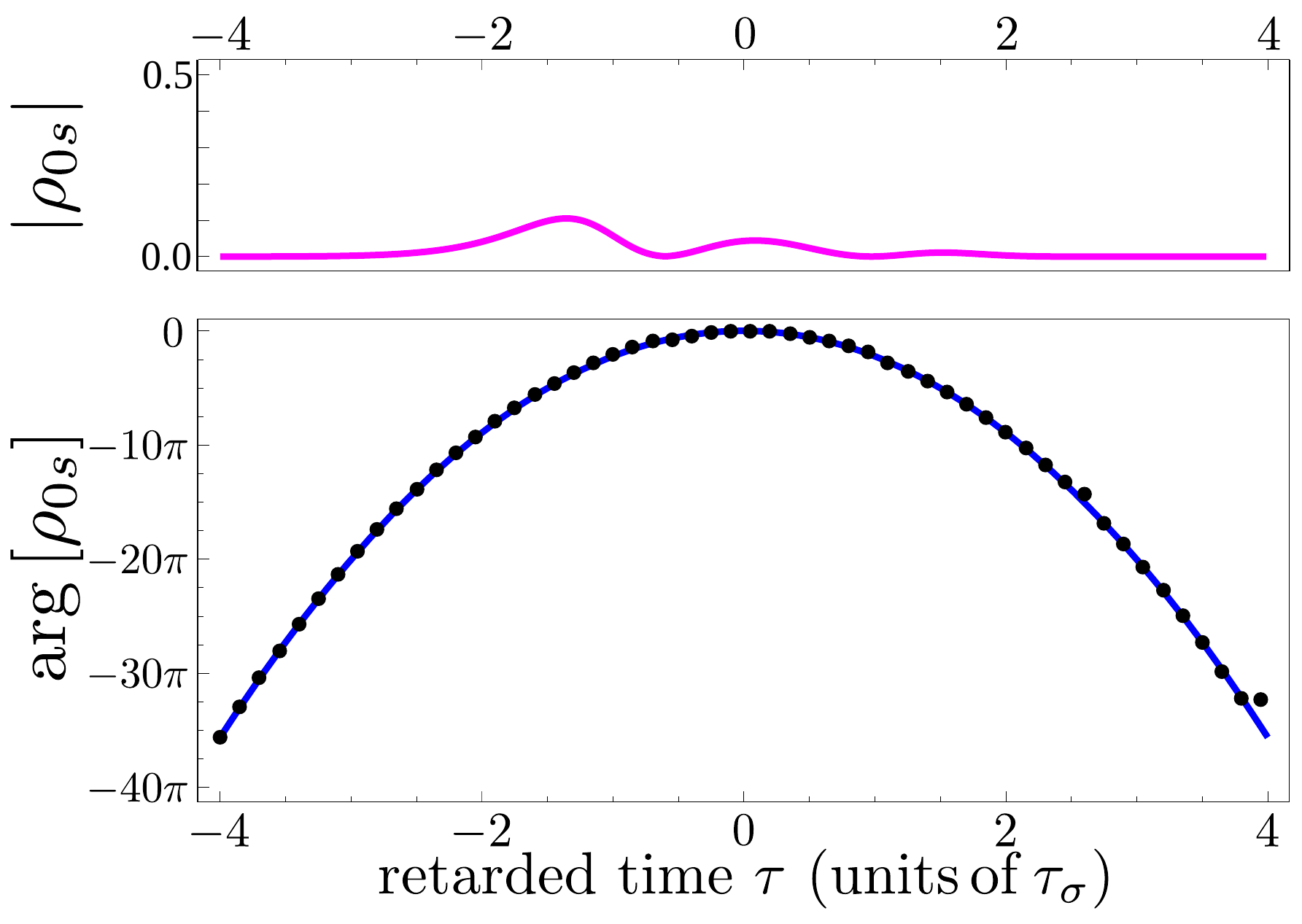}}}
\subfigure[\label{fig:coha}]{\resizebox{6.8 cm}{!}{\includegraphics*{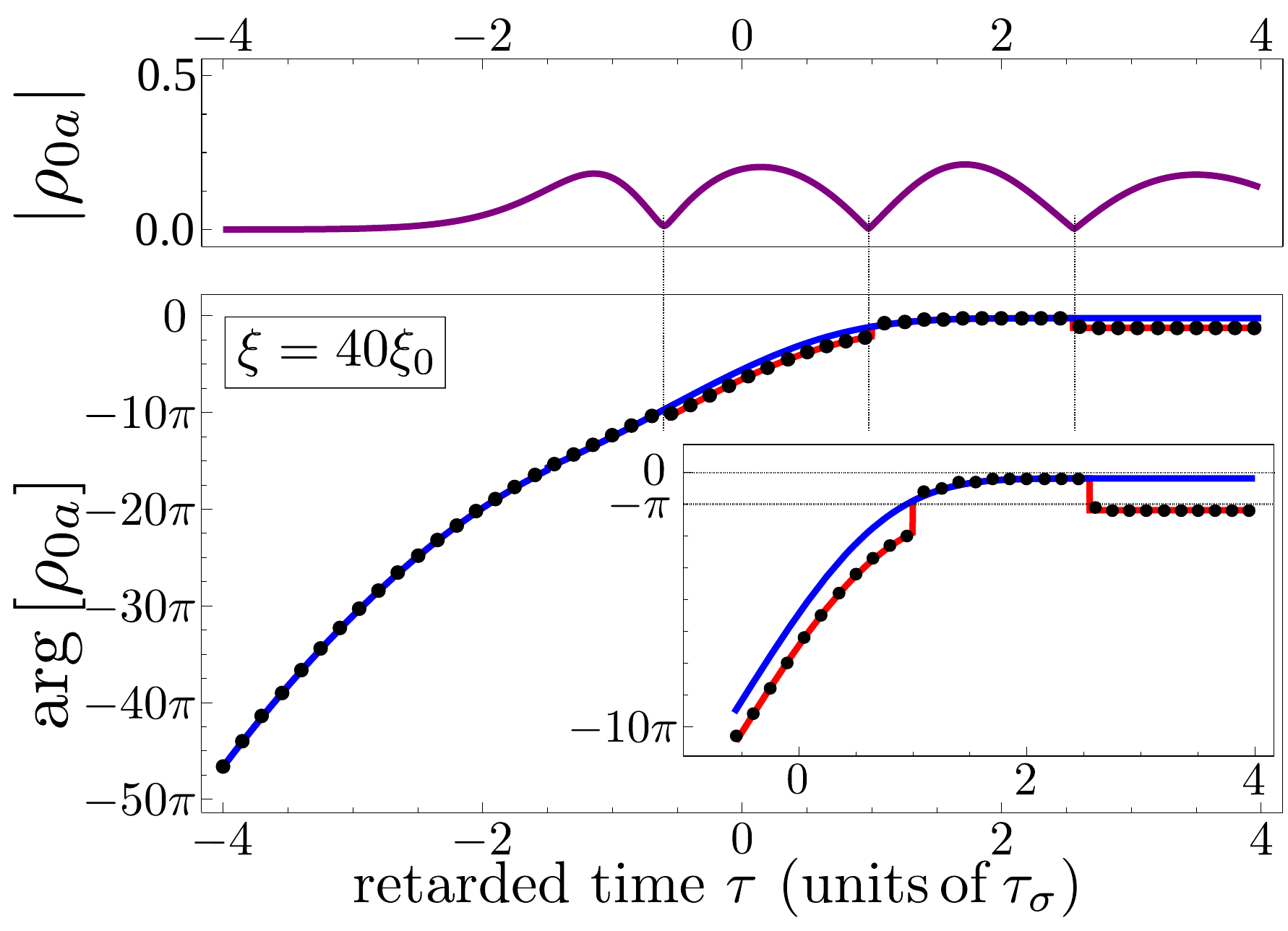}}}
\caption{(Color online) \textbf{Time evolution of the absolute values and phases of the atomic coherences $\rho_{0s}$ (a.) and $\rho_{0a}$ (b.) inside the medium ($\xi=40\xi_{0}$). (Inset: Magnification of the plot showing the time evolution of the phases between $\tau=-1$ and $\tau=4$.)} The solid lines in case of the absolute values and the dots in case of the phases are results of numerical calculation with the same parameters as in Fig.~\ref{fig:states0}. The solid lines  that connect the dots are functions fitted on the points. The same fitting functions were used as in case of the coherences at the boundary (see Fig.~\ref{fig:cohs0}) and the fitting parameters also proved to be the same. The only exception is $\beta_0$ which needs to be shifted by $\pi$  at the point where there is a jump in the data.}
\label{fig:coherences}
\end{figure}

The behavior of the phase functions of $\Omega_s$ and $\Omega_a$ is understandable if we look at the coherences between the excited state and the antisymmetric and symmetric states --- $\rho_{0a}\fun{\xi,\tau}$ and $\rho_{0s}\fun{\xi,\tau}$ --- for the atoms at a typical inner location of the medium. It can be seen in Fig.~\ref{fig:coherences} that although the absolute values of the coherences are smaller for atoms at $\xi\gg\xi_0$, the functions of their phases over time remain the same as it was at the boundary apart from phase-jumps of $\pi$ in the case of $\rho_{0a}\fun{\xi,\tau}$. Since the macroscopic polarizations of the medium for  the laser modes $\Omega_a$ and $\Omega_s$ are proportional to $\rho_{0a}$ and $\rho_{0s}$, respectively, the result for the phases of the laser modes is plausible.

\begin{figure}[!hbt]
\includegraphics[width=6.8cm, clip=true]{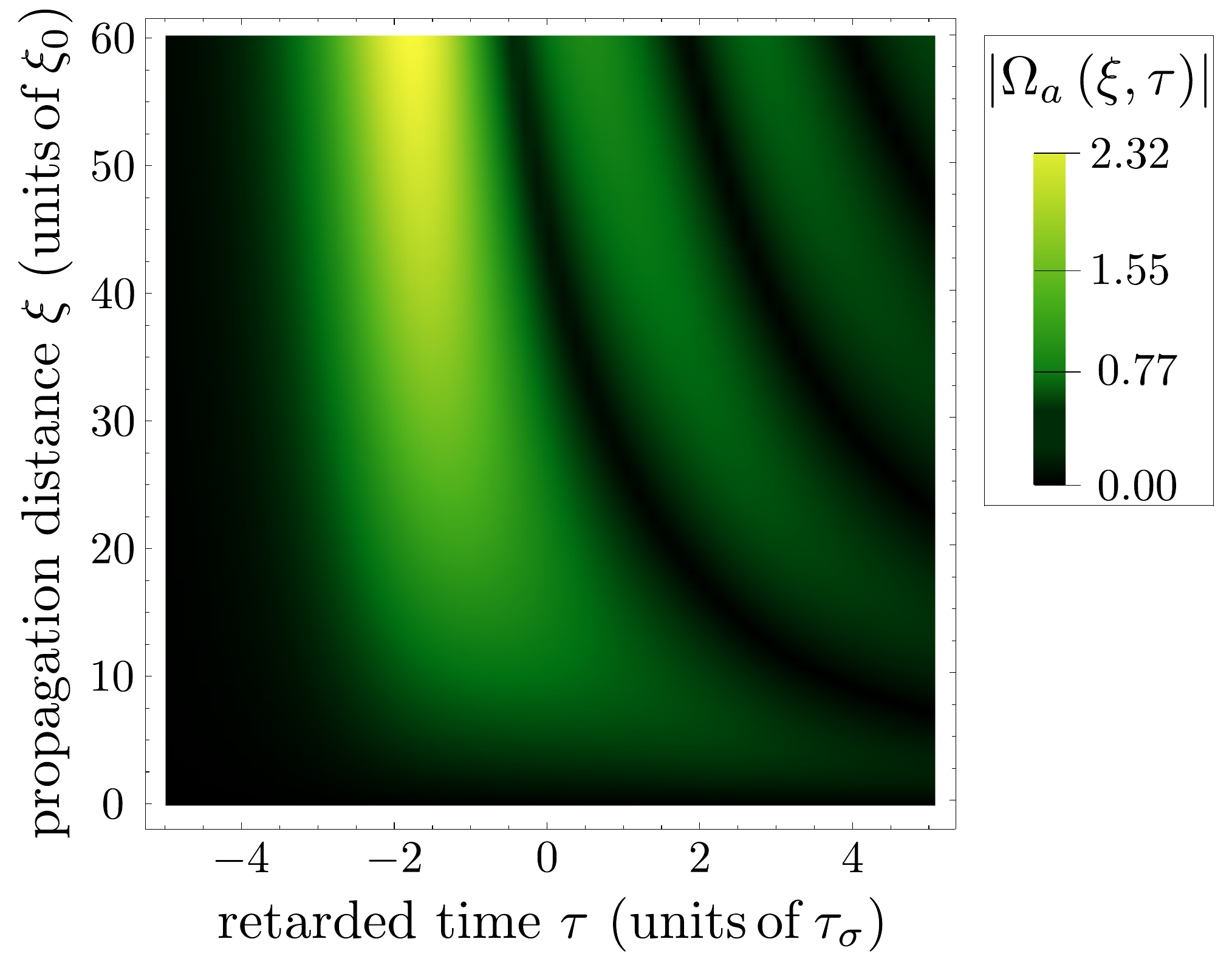}
\caption{(Color online) \textbf{Absolute value of the effective Rabi-frequency  $\Omega_a$ as a function of the retarded time coordinate and the propagation distance} The parameters used for the calculation are the same as in Fig.~\ref{fig:states0}.}
\label{fig:Omega}
\end{figure}

The phase-jumps of $\pi$ in the phase function of $\Omega_a\fun{\xi\gg\xi_0}$ and also of $\rho_{0a}\fun{\xi\gg\xi_0}$ occur when the quantities change there signs from positive to negative or the other way around. This oscillatory behavior of the antisymmetric mode $\Omega_a$ and the coherence between the excited state and the antisymmetric state $\ket{a}$ can be understood as follows. The time evolution of the phase of the antisymmetric mode is close to constant for approximately $\tau>1.5\tau_{\sigma}$. This means that the excited and the antisymmetric ground states $\ket{0}$ and $\ket{a}$ are coupled by a train of small resonant pulses having nearly constant frequency, with a phase difference of $\pi$ between the adjoining pulses. The pulse areas of these small pulses are much less than $\pi$, thus as a result of the interaction with one pulse, a small part of a Rabi-cycle proceeds between states $\ket{0}$ and $\ket{a}$. After this interaction, the process is reversed because of the next pulse which has an opposite sign ($\pi$ phase-shift). Once this oscillation appears in the coherence $\rho_{0a}$, it affects the formation of the field $\Omega_a$, that is why the further this pulse propagates the more small pulses appear (see Fig.~\ref{fig:Omega}).
 
To summarize, we found that --- along with the almost-unchanged chirped $\Omega_s$ symmetric mode --- an antisymmetric mode $\Omega_a$ appears  which is linearly chirped at the beginning and has a constant frequency at the end of the pulse.

\subsubsection{The time evolution of the atoms' state inside the medium}
\label{subsub:atoms}
\begin{figure}[!hbt]
\includegraphics[width=6.8cm, clip=true]{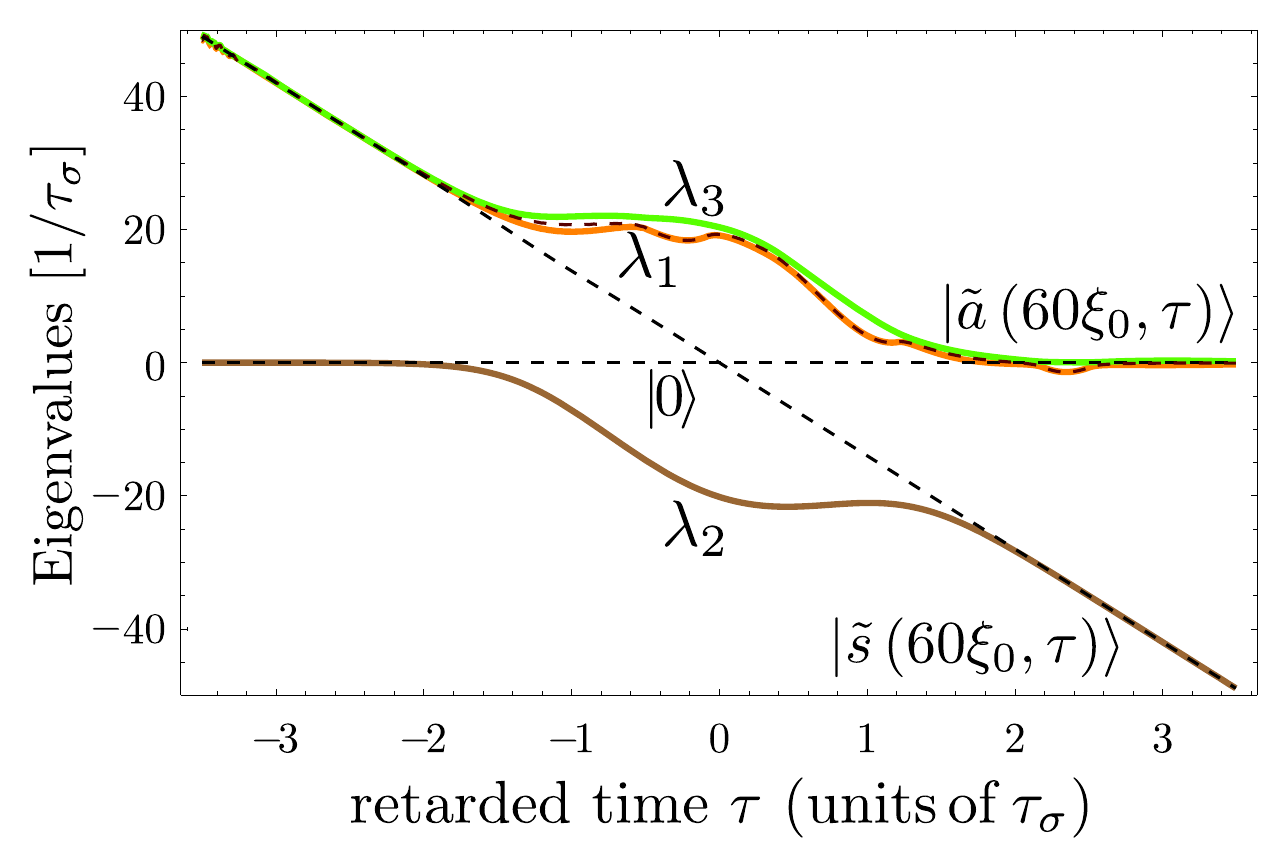}
\caption{(Color online) \textbf{Eigenvalues of the Hamiltonian $\hat{\mathcal{H}}_{sa}$ (Eq. \eqref{eqn:Hamrotsa}) which describes the atom-laser system far from the boundary of the medium ($\mathbf{\xi=40\xi_0}$).} The notations are the same as in Fig.~\ref{fig:eiv0}}
\label{fig:eiv}
\end{figure}
Let us analyze the time evolution of the state of the atoms caused by the interaction with the above-described two modes $\Omega_s$ and $\Omega_a$ at a given location inside the medium. First we use the adiabatic approximation. The eigenvalues of the Hamiltonian $\mathcal{H}_{sa}\fun{\xi,\tau}$ defined in Eq.~\eqref{eqn:Hamrotsa} are presented in Fig.~\ref{fig:eiv}.

There is an obvious difference between this energy spectrum and the one of the interaction Hamiltonian characteristic to the atoms in the boundary (cf. Fig.~\ref{fig:eiv0}). Namely, the eigenenergy of the diabatic state $\ket{a}$ has the constant value of zero and since $\ket{a}$ is completely uncoupled at the boundary --- $\Omega_a\fun{\xi=0,\tau}=0$ ---, $\ket{a}$ is an adiabatic eigenstate as well, with a zero eigenvalue constant in time. In contrast, for atoms at inner locations $\xi\gg\xi_0$ of the medium, $\Omega_a\fun{\xi\gg\xi_0,\tau}$ is, though small but nonzero, and as it is discussed in the previous section, has a nontrivial phase function over time. Since the time derivative of this phase is incorporated into the Hamiltonian $\hat{\mathcal{H}}_{sa}\fun{\xi\gg\xi_0,\tau}$ in the interaction picture defined  by the rotating basis~\eqref{eqn:rotbasis}, the diabatic state $\ket{a}$ here has a nontrivial time-dependent eigenvalue. 

At the beginning of the interaction, the atoms are prepared in the superposition of the antisymmetric and symmetric diabatic states $\ket{a}$ and $\ket{s}$. As it can be seen in Fig.~\ref{fig:eiv}, the eigenvalues of these adiabatic states correspond to the adiabatic eigenvalues $\lambda_1$ and $\lambda_3$. Similar to the case at the boundary, the population, which was initially in state $\ket{a}$ remains there since $\lambda_1$ perfectly overlaps the diabatic energy of the antisymmetric state. On the other hand, the adiabatic eigenvalue $\lambda_3$ moves further from the diabatic energy of state $\ket{s}$ and remains close to the energy of $\ket{a}$. Based on the adiabatic theorem~\cite{Oreg1984, Grischkowsky1976} we expect the population, that was initially in the symmetric state $\ket{s}$ to be transferred into the antisymmetric state $\ket{a}$. 

We also have to notice that these eigenstates --- $\lambda_1$ and $\lambda_3$ --- move close to the diabatic energy of the excited state $\ket{0}$ at the end of the interaction, for $\tau>1.5\tau_{\sigma}$. We do not expect a large excitation of the atoms, however. Comparing Figs.~\ref{fig:eiv} and~\ref{fig:Rabia}, one realizes that the coupling $\Omega_a$ is very small for $\tau>1.5\tau_{\sigma}$.

\begin{figure}[!hbt] 
\includegraphics[width=6.8cm, clip=true]{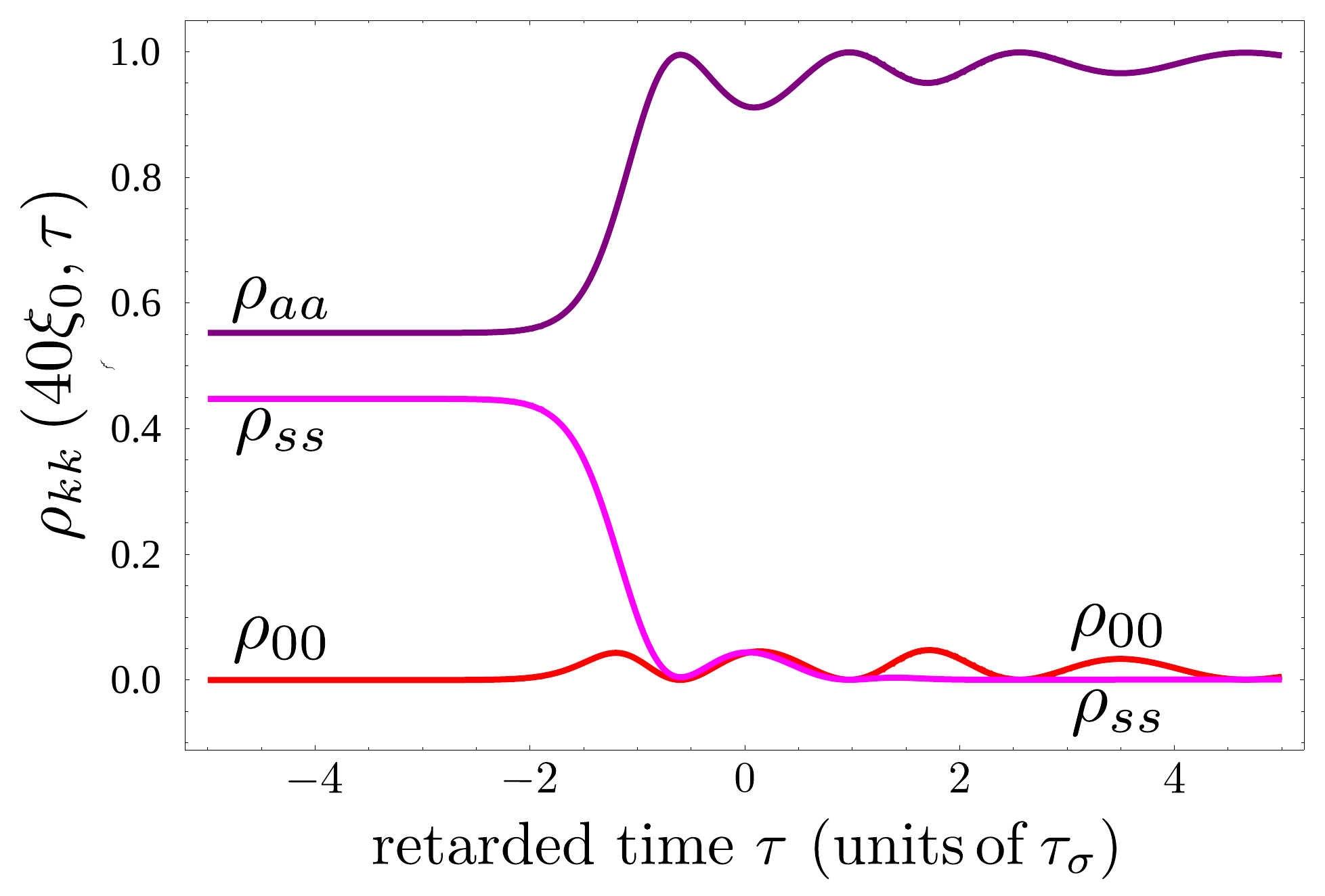}
\caption{(Color online) \textbf{Dynamics of the populations of the atomic states inside the medium ($\xi=40\xi_0$) in the symmetric-antisymmetric  basis.} The parameters used for the calculation are the same as given in Fig.~\ref{fig:states0}. }
\label{fig:pop161}
\end{figure}

The results of our numerical calculations --- presented in Fig~\ref{fig:pop161} --- fit perfectly to the considerations in the adiabatic picture. That is, the majority of the population is transferred into the antisymmetric state $\ket{a}$, and the excitation of the atom is drastically reduced (it does not exceed 5\% during the whole interaction). Thus, although complete population trapping is not established as in the case of the constant frequency matched pulses~\cite{Harris1995}, a quasi-dark state is created by the two modes $\Omega_s$ and $\Omega_a$.

\begin{figure}[!hbt]
\includegraphics[width=6.8cm, clip=true]{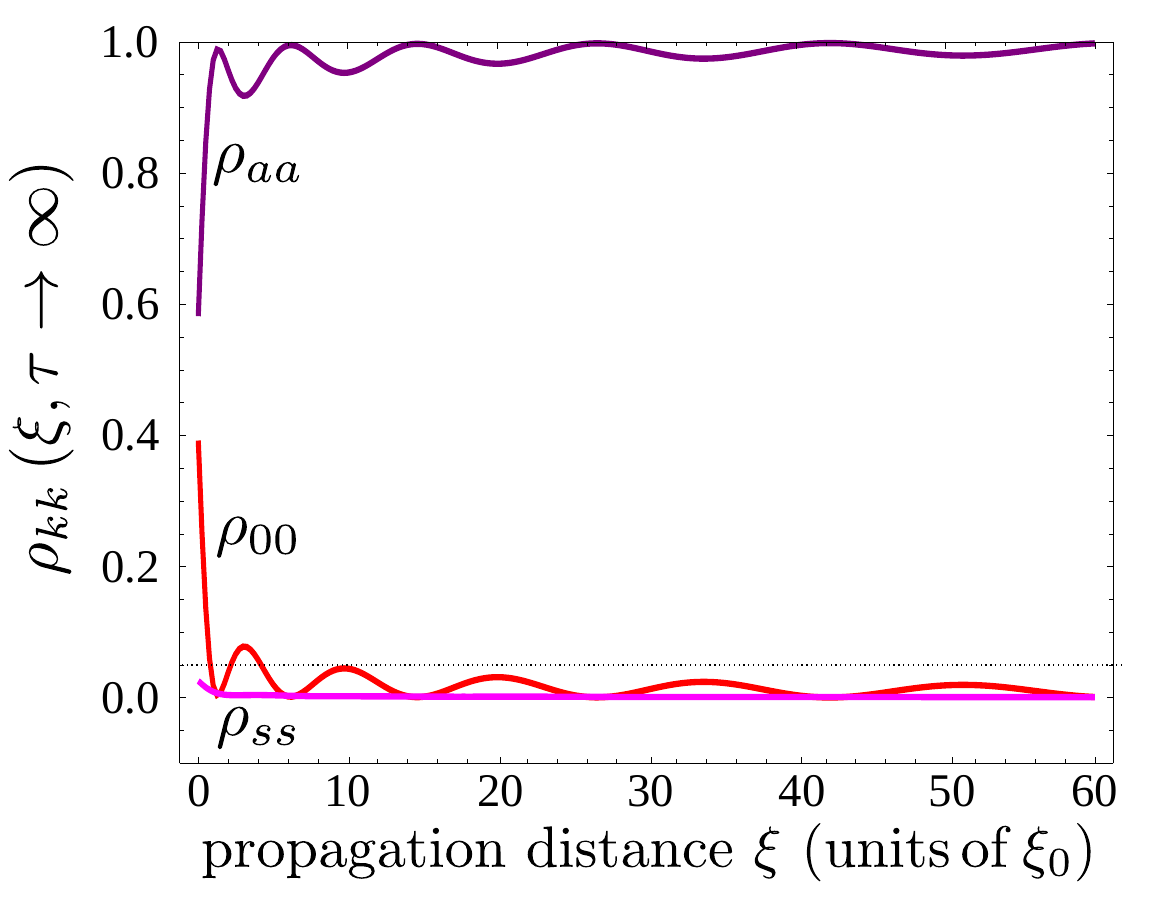}
\caption{(Color online) \textbf{Final populations of the atoms at different space points $\xi$ in the symmetric-antisymmetric basis.} After a few absorption length $\xi_0$ of propagation, the pulse pair transfers the majority of the atomic population to the antisymmetric superposition of the ground states. The same parameters were used for the calculation as in Fig.~\ref{fig:states0}.}
\label{fig:finpopsb}
\end{figure}

It is easy to see how the population control process induced by the laser pulse pair changes in course of their propagation in the medium by looking at Fig.~\ref{fig:finpopsb}. This figures show the populations of the states $\ket{a}$, $\ket{s}$ and $\ket{a}$ after the atoms' interaction with the laser pulses $\Omega_s$ and $\Omega_a$. At the boundary and within one absorption length $\xi_0$, a significant part of the population is transferred into the excited state. After a few absorption lengths, the excitation of the atom decreases, and although shows an oscillatory behavior, it is always below $5\%$ for atoms that are located at space positions $\xi>6 \xi_0$.

We would like to point out here that the self-organization mechanism of the chirped pulse pair is based on a population transfer mechanism inside the medium substantially different from other cases found in literature (see e.g.~\cite{Harris1995, Kozlov2009} with references therein) dealing with propagation of constant frequency pulses. Both of the latter are based on the fractional-STIRAP mechanism,  where the final distribution of the population among the ground states is determined by the intensities of trailing edges of laser pulses~\cite{Kozlov2009}. In case of the matched propagation of two constant-frequency pulses~\cite{Harris1995, Harris1994, Harris1997}, the originally simultaneous laser pulses become shifted in time with respect to each other, allowing for a lossless propagation. Since there can oscillations occur on the pulses' envelope, the final state can be significantly different for atoms at different locations. In~\cite{Kozlov2009} this problem is overcome by using adiabatons as boundary condition, but it has the disadvantage of the need of precise control of the ingoing pulse shape.

In our case however, the adiabatic transfer occurs due to a different mechanism. This mechanism is based on the special time function of the antisymmetric mode $\Omega_a$, which is generated by the medium. The adiabatic transfer results in the complete emptying of the  symmetric state $\ket{s}$, and in the transfer of the majority of the population into the antisymmetric state $\ket{a}$.

\subsection{Description of the system in the original basis}
\label{sec:orig}
\begin{figure}[!hbt]
\subfigure[\label{fig:popch}]{\resizebox{5.3cm}{!}{\includegraphics*{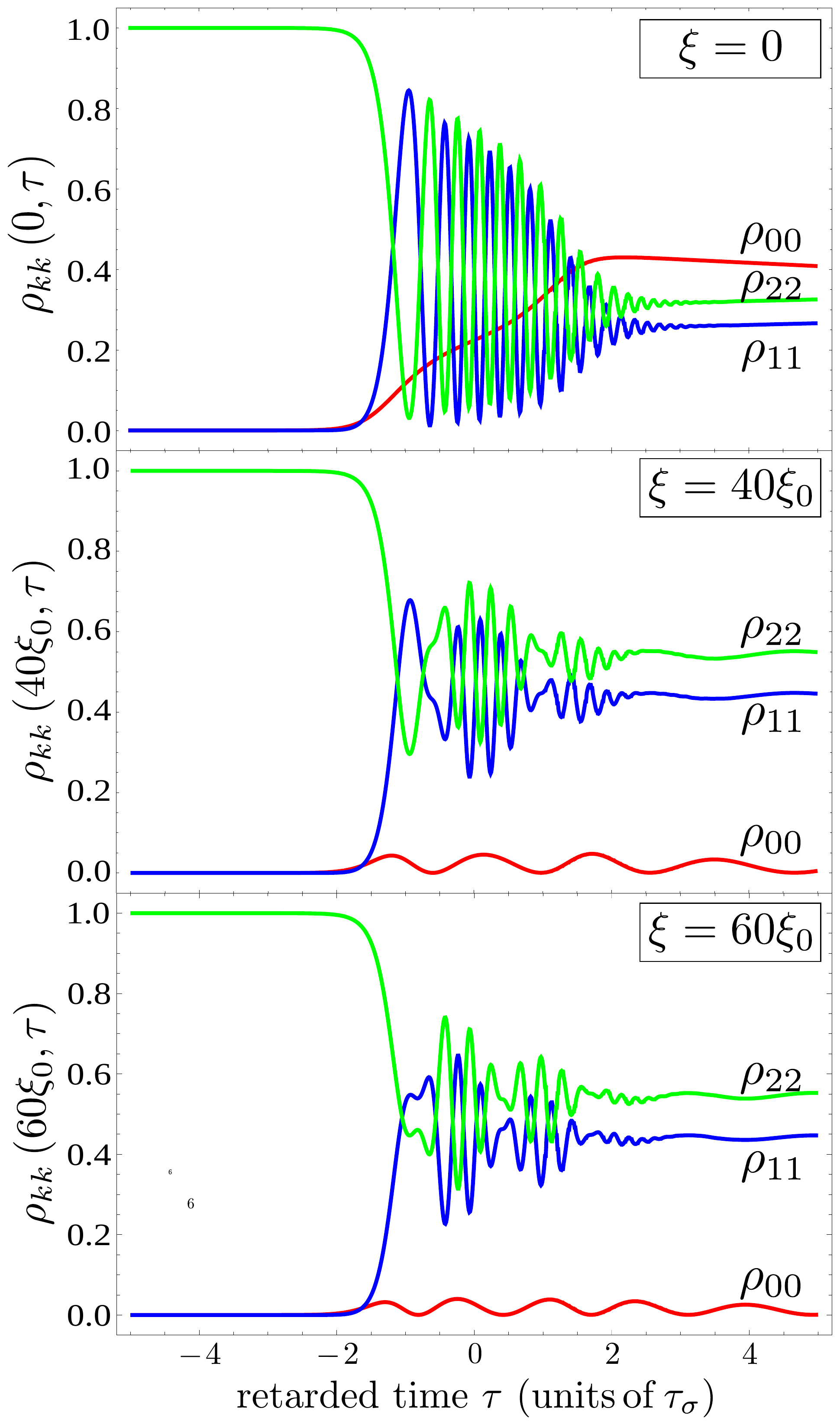}}}
\subfigure[\label{fig:popconst}]{\resizebox{5.3cm}{!}{\includegraphics*{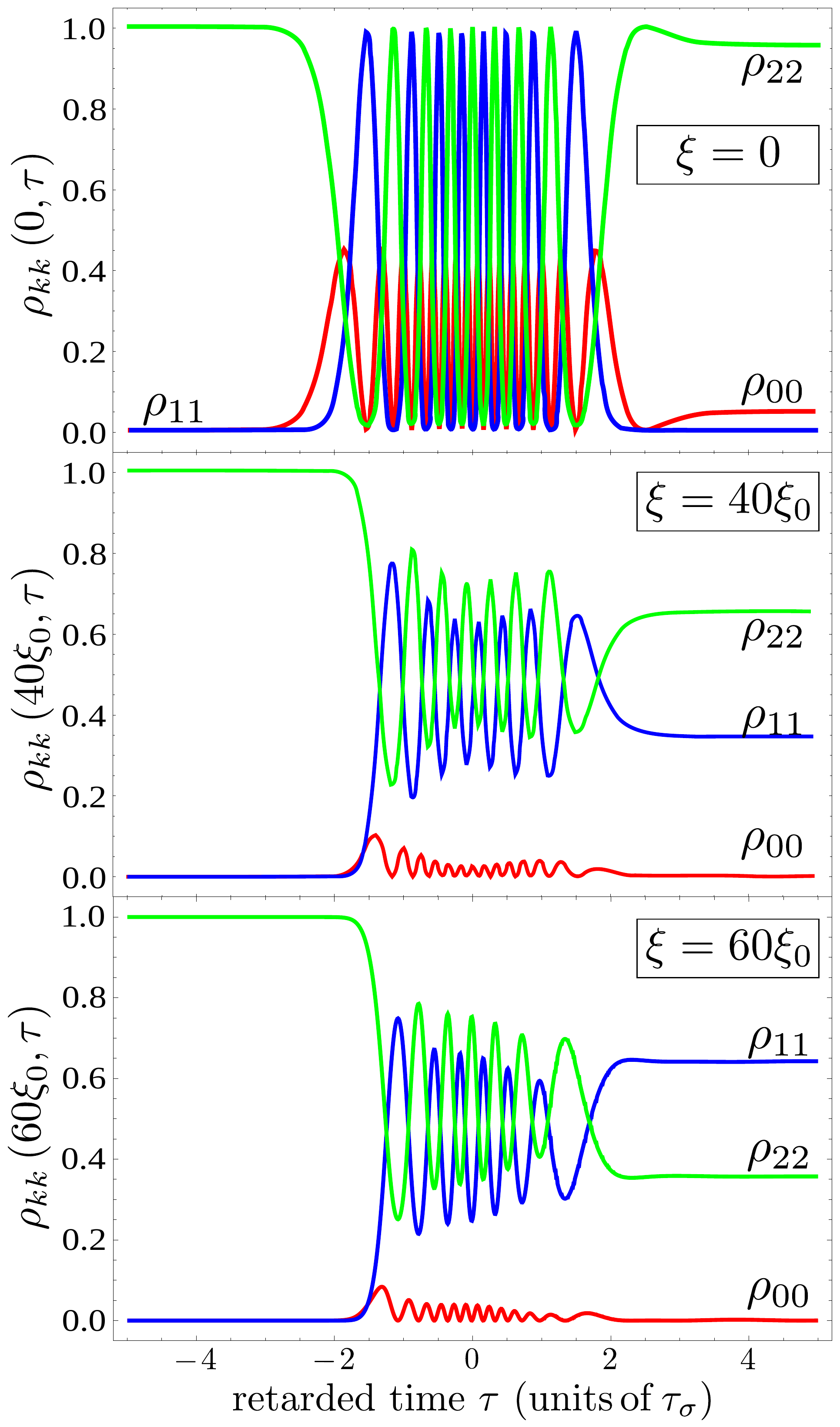}}}
\caption{(Color online) \textbf{Evolution of the populations as a function of the retarded time $\tau$ induced by a pair of a.) chirped (FC) and b.) constant-frequency pulses at the boundary and  at two typical propagation lengths inside the medium ($\xi=40\xi_0$ and $\xi=60\xi_0$).} The parameters used for the numerical calculations are the same as in Fig.\ref{fig:states0}.}
\label{fig:orpops}
\end{figure}

From the point of view of possible applications it is important to "translate" our results in the symmetric-antisymmetric basis and with the effective couplings $\Omega_a$ and $\Omega_s$ into the original atomic basis. Fig.~\ref{fig:orpops} presents the population transfer process induced by a pair of FC and constant-frequency pulses, respectively,  in the original atomic basis at the boundary and at two given propagation lengths inside the medium ($\xi=40\xi_0$ and $\xi=60\xi_0$). The dynamics of the populations induced by the FC (Fig.~\ref{fig:popch}) and the constant-frequency pulse pair shows a conspicuous difference. This pronounced difference demonstrates well the different underlying physics, explained in details in Subsection.~\ref{subsub:atoms} by using the symmetric-antisymmetric basis.

On one hand,at the boundary, the Raman-resonant FC pulse pair induces a coherent population transfer which distributes the population initially prepared in state $\ket{2}$ into a superposition of the three atomic states. The population of the excited state as a result of the interaction with the pulse pair is $\rho_{00}=\vartheta_2^2/\vartheta^2$ coinciding with the population of state $\ket{s}$ at the beginning of the interaction, while the population that remains in the ground states after the population transition process is $\rho_{11}=\fun{\vartheta_1^2\vartheta_2^2}/\vartheta^2$ and $\rho_{22}=\vartheta_1^4/\vartheta^2$, respectively (which coincides with the initial population of the antisymmetric state $\ket{a}$). 

Note that the requirement of the Raman-resonance between the couplings plays an important role in the whole process under consideration. That is, a substantially different population evolution is induced by Raman-detuned couplings of the two allowed transitions~\cite{Djotyan2004}, which leads to population transfer between the ground states, along with negligible excitation of the atom. This mechanism is rather similar for both a single atom and the atoms in an optically thick medium taking into account propagation of the interacting laser pulses.

\begin{figure}[!hbt]
\subfigure[\label{fig:finpopa}]{\resizebox{6.8cm}{!}{\includegraphics*{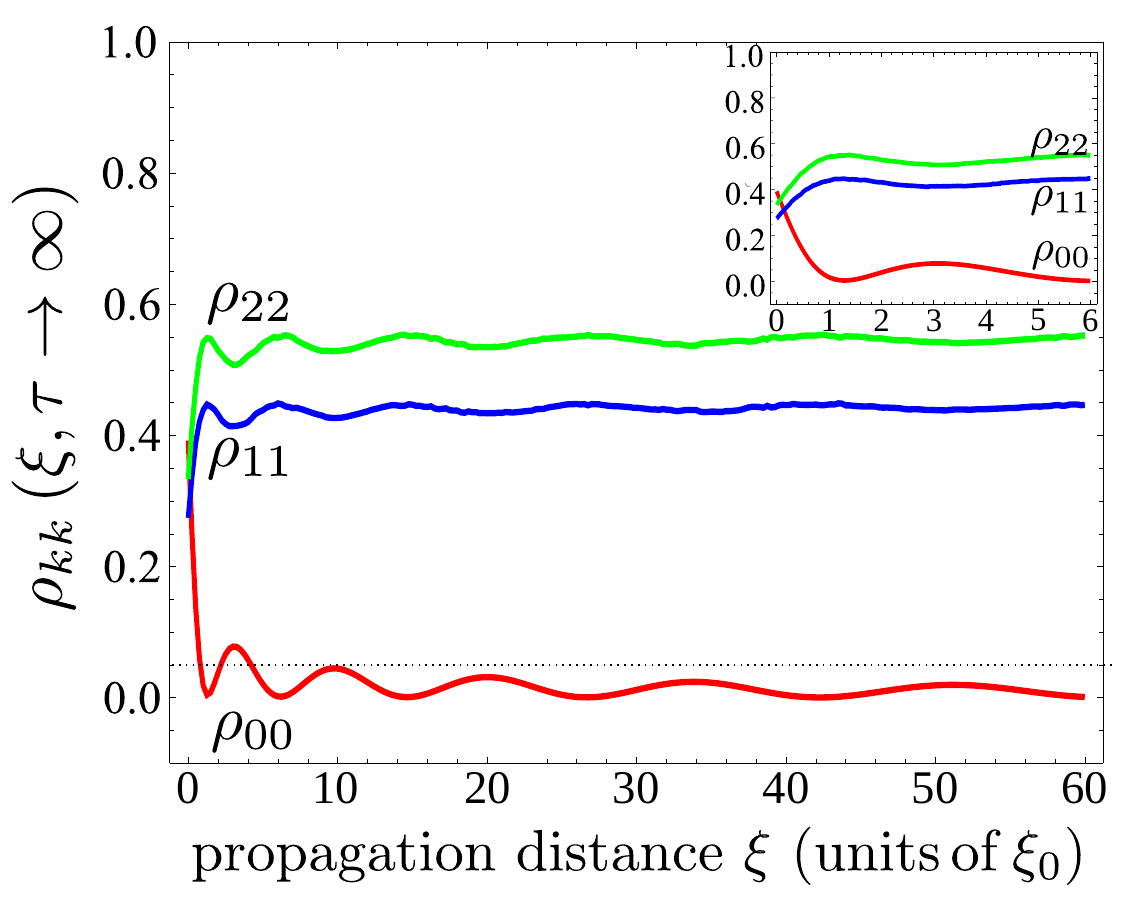}}}
\subfigure[\label{fig:finpopc}]{\resizebox{6.8cm}{!}{\includegraphics*{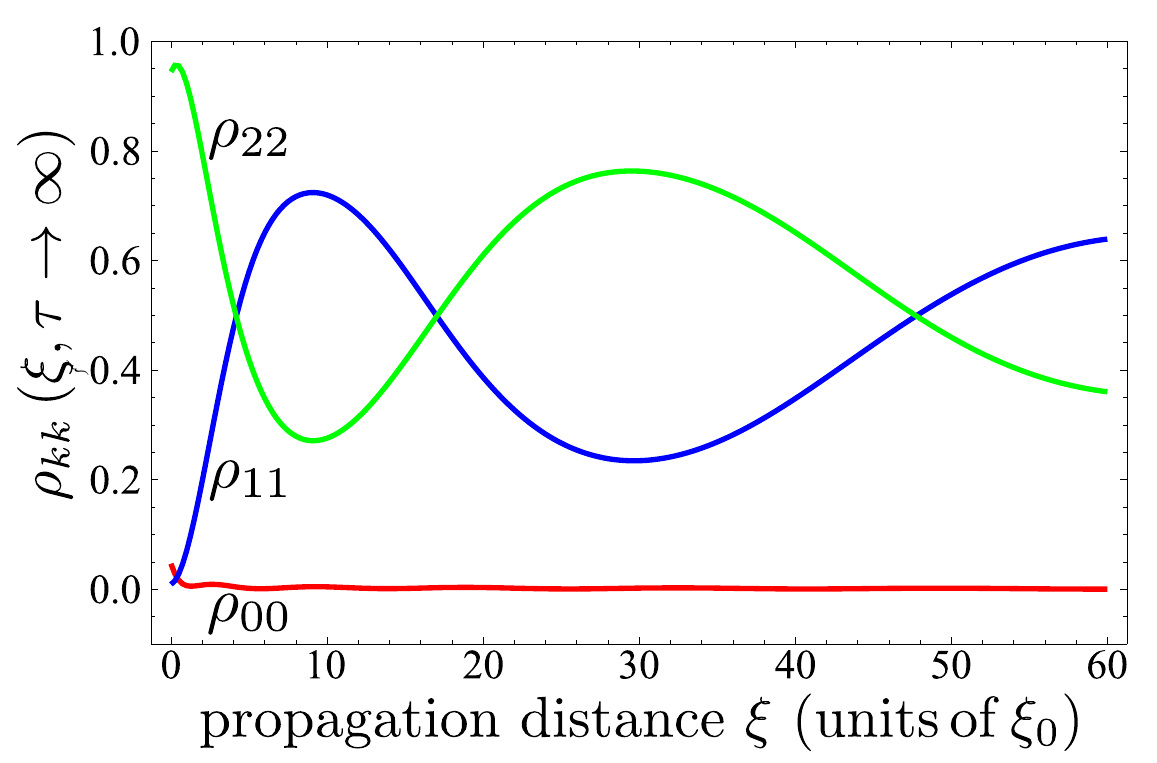}}}
\caption{(Color online) \textbf{Final populations of the atoms at different space points $\xi$ in case of a.) frequency-chirped b.) constant frequency pulse pair. Inset: Final populations close to the boundary ($\xi\leq 6\xi_0$)} for a FC pulse pair. After a few absorption length $\xi_0$ of propagation, the FC pulse pair transfers the majority of the atomic population to the antisymmetric superposition of the ground states, while in case of the matched pulses having constant frequency, the final state strongly varies with $\xi$. The parameters used for calculations are: $\vartheta_1=15 \ival{1/{\tau_{\sigma}}}$, $\vartheta_2=13.5\ival{1/{\tau_{\sigma}}}$, $\beta=7\ival{1/{\tau_{\sigma}^2}}$ and $\beta=0$, respectively.}
\label{fig:finpops}
\end{figure}

In contrast, in the present case of Raman-resonant coupling,  both the pulse envelopes and the time function of the phases of the pulses are modified by the interaction  with the medium after a short propagation length. The modification takes place in such a way that instead of exciting the atom, the pulses drive the main part of the population into the  $\fun{\vartheta_2\ket{1}-\vartheta_1\ket{2}}/\vartheta$ superposition of the ground states (see Fig.~\ref{fig:finpops}).

On the other hand, the constant-frequency pulses induce a Rabi-oscillation between the excited state and the superposition of states $\ket{1}$ and $\ket{2}$, which corresponds to the antisymmetric state $\ket{a}$. Similarly to the FC-case, the constant-frequency pulses become `matched'~\cite{Harris1995} (cf. ~\ref{subsub:atoms}), that is the excitation of the atoms is drastically reduced inside the medium.

Note that, neither in the FC, nor in the constant-frequency case is all the population transferred to the antisymmetric superposition. In case of the FC pulse pair it is the excited state which is slightly populated but in the weakly decaying regime under consideration this does not disturb significantly the preparation of the medium into a well-defined superposition state controlled by the peak Rabi frequencies of the ingoing pulses, as is shown in Fig.~\ref{fig:finpopa}. Fig.~\ref{fig:finpopc} demonstrates the advantage of using FC pulses. Since the matched pulses having constant frequencies transfer different (though small) amounts of population into the symmetric superposition $\ket{s}$ in atoms at different locations $\xi$, the induced population distribution among the atomic states changes significantly as a function of $\xi$ (see Fig.~\ref{fig:popconst} and~\ref{fig:finpopc}).
We would like to emphasize that the difference in the state of the medium created by the chirped and the constant-frequency pulse pair is due to the different underlying physics, explained in details in Subsection.~\ref{subsub:atoms}.

\begin{figure}[!hbt]
\subfigure[\label{fig:pulsesa}]{\resizebox{6.8cm}{!}{\includegraphics*{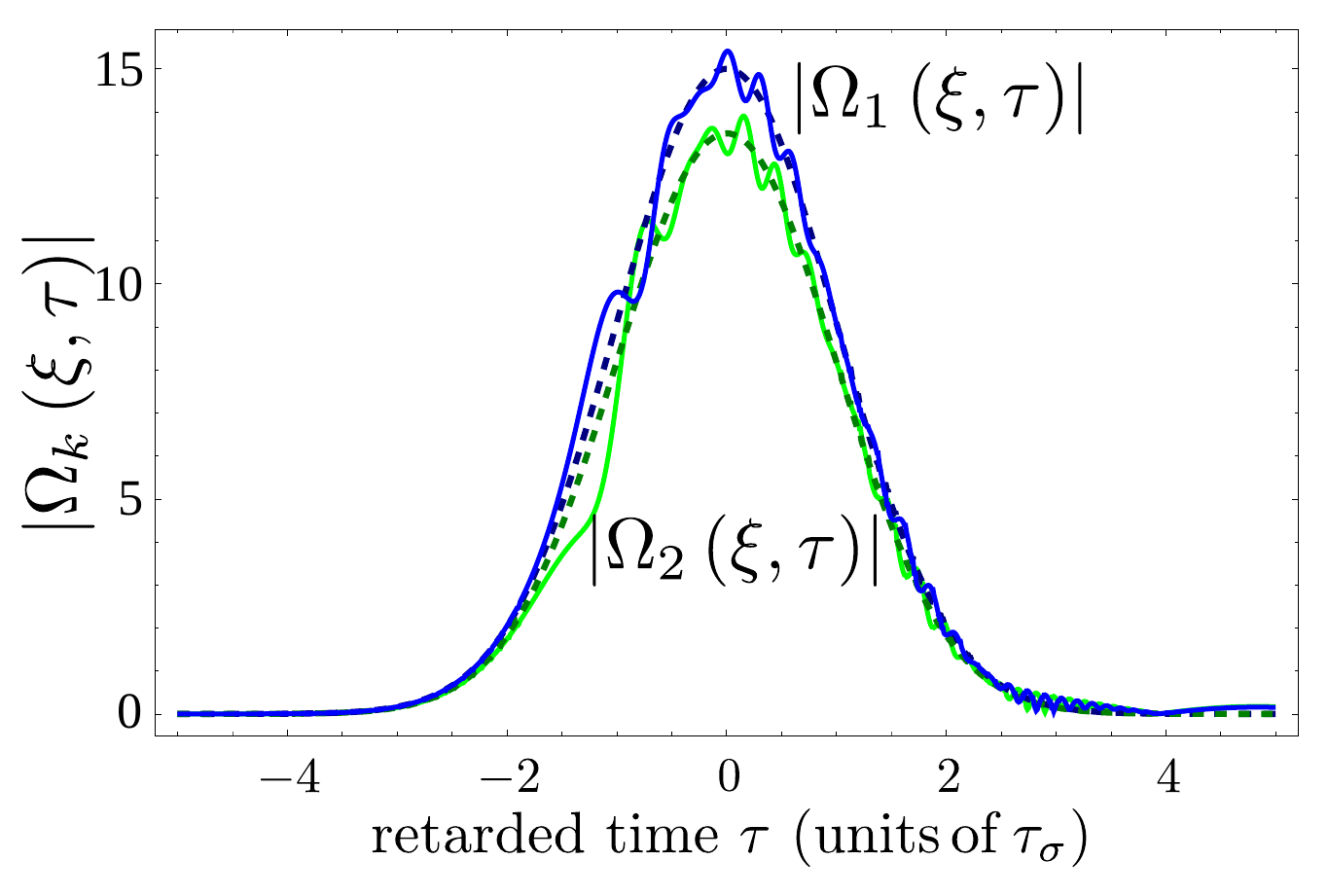}}}
\subfigure[\label{fig:pulsesarg}]{\resizebox{6.8cm}{!}{\includegraphics*{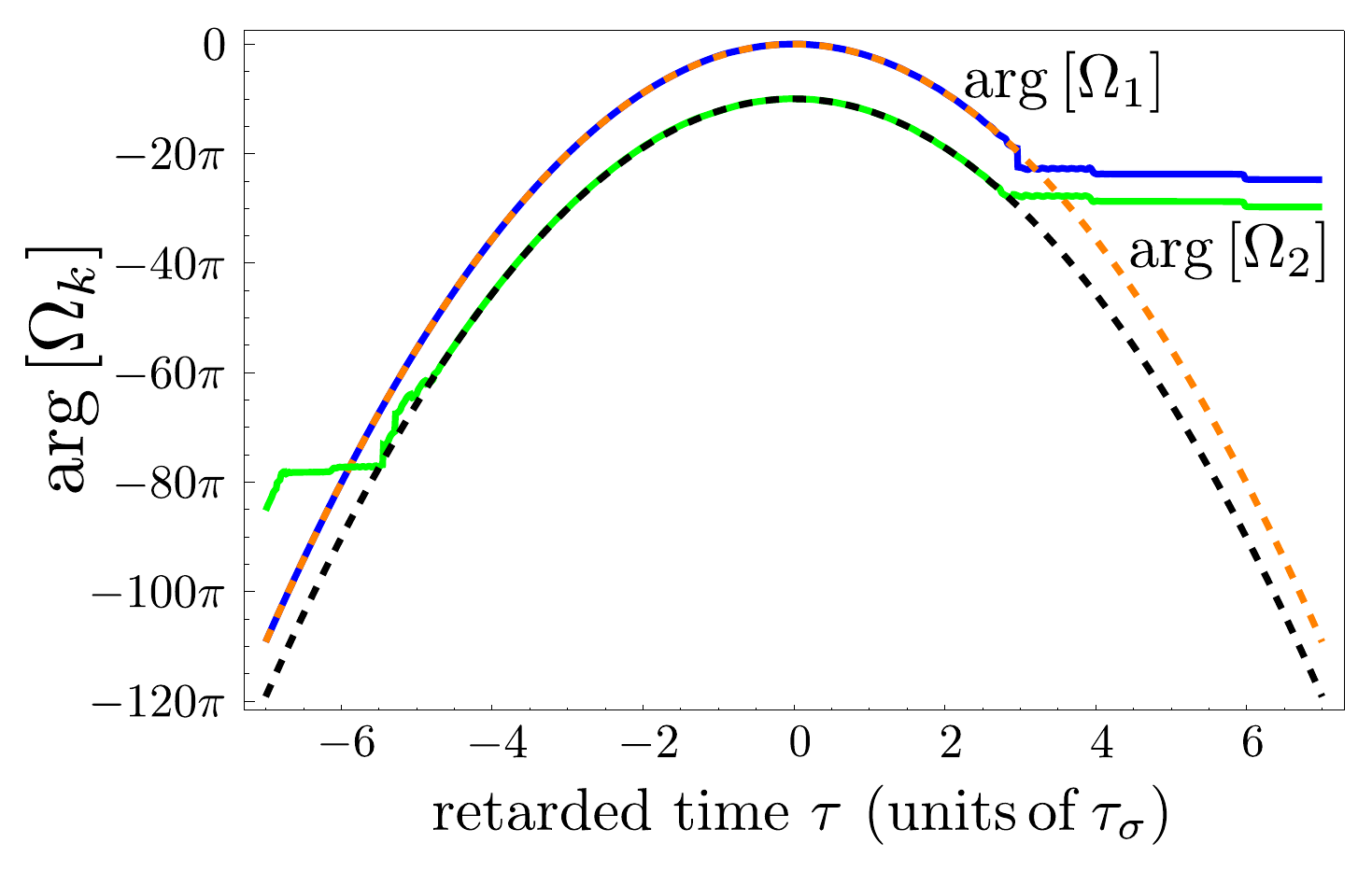}}}
\caption{(Color online) \textbf{a.) Envelope functions of the pulses at the boundary (dashed lines) and at $\xi=40\xi_0$ (solid lines) b.) The phases of the Rabi-frequencies as a function of the retarded time $\tau$ at the boundary (dashed lines) and inside the medium (solid lines)} The same parameters was used for numerical calculation as given in the caption of Fig.\ref{fig:states0}.}
\label{fig:pulses}
\end{figure}

The Rabi-frequencies can be expressed by the symmetric and antisymmetric Rabi-frequencies as
\begin{subequations}
\begin{align}
\Omega_1=\frac{\vartheta_1}{\vartheta}\Omega_s+\frac{\vartheta_2}{\vartheta}\Omega_a\\
\Omega_2=\frac{\vartheta_2}{\vartheta}\Omega_s-\frac{\vartheta_1}{\vartheta}\Omega_a.
\end{align}
\end{subequations}
As the pulses propagate in the medium, energy is transferred from $\Omega_2$ (which couples the transition where the atoms are prepared) to $\Omega_1$. Since $\Omega_a$ is one order of magnitude weaker than $\Omega_s$, even after propagation length of many times $\xi_0$ (c.f.~\ref{fig:Rabi}), the distortion is small. In this sense it can be stated that the FC pulse pair propagates quasi-transparently and that it can prepare a well-defined coherent superposition of the ground states in an extended medium of relatively large optical depth (see Fig.~\ref{fig:pulses}).

\section{Summary}
\label{sec:sum}
We have analyzed the propagation of a pair of Raman-resonant, linearly frequency modulated strong laser pulses in an optically thick medium, which is modeled as a motionless and noninteracting ensemble of $\Lambda$-atoms. We have demonstrated that quasi-lossless propagation of FC pulses is possible not only when the medium is initially prepared in a quasi-dark state~\cite{Demeter2007}, but through a \emph{matching} effect between the two pulses. Namely, although the Raman-resonant pulse pair causes a significant excitation in the atoms close to the boundary of the medium, the excitation of the atoms becomes negligible in the medium at larger propagation length. Excitation of the atoms near the boundary of the medium, however, plays an important role in generation of a macroscopic polarization, which interaction with the FC pulses results in the matched quasi-lossless propagation of the pulses in the optically thick medium. 

By analyzing the dressed states of the atoms, we have demonstrated that the FC pulse pair induces a population transfer mechanism substantially different from the transfer process typical for the matched pulses having constant carrier frequency. The FC pulse pair, in course of its propagation transfers the majority of the atoms of the medium into approximately the same coherent superposition of their ground states. In contrary, the population distribution among the ground states induced by the constant frequency pulse pair may change significantly at different locations in the medium. 

 We have shown that the composition of the coherent superposition, established by the propagating FC pulse pair, depends on the peak amplitudes of the these laser pulses at the boundary of the medium. Therefore, the magnitude of the coherence created by the interaction may be tuned by parameters which are easily controllable experimentally.

The obtained results, especially those concerning the robust creation of coherence between atomic metastable (ground)  states in a spatially extended, optically thick medium may find important applications in schemes of frequency conversion through nonlinear optical mixing processes, as well as in other nonlinear processes where the initial preparation of an extended medium in a coherent superposition state is needed~\cite{Scully1992191,Scully1992, Mompart2000,Zibrov1996, Lukin1999, Kolesov2006, Sautenkov2000, Wojciechowski2010}. 

Another possible application of the process may be in the realm of quantum communication, where chirped pulses have been proven useful in photon-echo based quantum memories~\cite{Damon2011}. For this direction of usage, the analysis of the propagation mechanism in the presence of inhomogeneous broadening is needed, which we will be the subject of our next article.

\section*{Acknowledgements}
This work was funded by the Research Fund of the Hungarian Academy of Sciences (OTKA)
under contract  NN 78112; the ELI- 09-1-2010-0010 grant; and T\'AMOP 4.2.4.A/2-11-1-2012-0001 `National Excellence Program'.

%\newpage
%\bibliography{2}{}
%

\end{document}